\documentstyle[aps,prl,preprint,floats,epsfig,mytab]{revtex}

\def\displaytwo#1#2#3#4#5{ 
   \begin{figure}[#1]
   \centerline{\psfig{figure=#2.ps,height=3.4in,clip=}} 
   \Caption{#4}\vskip 1.0truecm
   \label{#2}
   \centerline{\psfig{figure=#3.ps,height=3.4in,clip=}}
   \Caption{#5}
   \label{#3}
   \end{figure}}
\newcommand{\psfiletwoBB}[5]{ 
  \begin{minipage}{\linewidth}
    \parbox[b]{.49\linewidth}{%
      \begin{center}
        \setlength{\epsfxsize}{#5\linewidth}\leavevmode\epsfbox[#1]{#2}
     \end{center}
    }
    \hfill
    \parbox[b]{.49\linewidth}{%
      \begin{center}
        \setlength{\epsfxsize}{#5\linewidth}\leavevmode\epsfbox[#3]{#4}
     \end{center}
    }
  \end{minipage}
}

\def\display4#1#2#3#4#5#6#7#8{
   \begin{figure}[#1]
   \begin{center}
   \hbox{
   \quad\hskip-1cm
   \parbox[t]{#7in}{ \psfig{figure=#3.ps,height=#7in} }
   \parbox[t]{#7in}{ \psfig{figure=#4.ps,height=#7in} }
   }
   \hbox{
   \quad\hskip-1cm
   \parbox[t]{#7in}{ \psfig{figure=#5.ps,height=#7in} }
   \parbox[t]{#7in}{ \psfig{figure=#6.ps,height=#7in} }
   }
   \vskip0.5cm\Caption{#8}
   \label{#2}
   \end{center}
   \end{figure} }

\def\dbline{\noalign{\vskip 0.15truecm\hrule}\noalign{\vskip 2pt}\noalign{\hrule \vskip 0.15truecm}}
\def\sgline{\noalign{\hrule}}
\def\piz{\pi^0}
\def\ra{\rightarrow}
\def\etal{{\it et al.}}
\newcommand{\xs}{\mbox{$X_S$}}

\newcommand{\DE}{\mbox{$\Delta{\rm E}$}}

\newcommand{\xf}{\mbox{${\cal F}$}}
\newcommand{\hel}{\mbox{${\cal H}$}}

\newcommand{\gaga}{{\gamma\gamma}}

\newcommand{\omegappp}{\mbox{$\omega\ra\pi^+\pi^-\piz$}}
\newcommand{\etagamgam}{\mbox{$\eta\ra\gaga$}}
\newcommand{\etathreepi}{\mbox{$\eta\ra\pi^+\pi^-\pi^0$}}
\newcommand{\etagg}{\mbox{$\eta_{\gaga}$}}
\newcommand{\kshpp}{\mbox{$K_S\ra\pi^+\pi^-$}}
\newcommand{\etatogg}{\mbox{$\eta\ra\gaga$}}

\newcommand{\etaprepp}{\mbox{$\etapr\ra\eta\pi^+\pi^-$}}
\newcommand{\etaprrg}{\mbox{$\etapr\ra\rho^0\gamma$}}

\newcommand{\calB}{\mbox{$\cal B$}}
\newcommand{\calL}{\mbox{$\cal L$}}

\newcommand{\Bkpi}{\mbox{$B^0\rightarrow K^+\pi^-$}}
\newcommand{\Bpipi}{\mbox{$B^0\rightarrow\pi^+\pi^-$}}

\newcommand{\etapr}{{\eta^{\prime}}}
\newcommand{\etaprk}{\mbox{$\etapr K$}}

\newcommand{\etaprkpd}{\mbox{$\etapr_{\eta\pi\pi}K^+$}}
\newcommand{\etaprkprg}{\mbox{$\etapr_{\rho\gamma}K^+$}}
\newcommand{\etaprkpfv}{\mbox{$\etapr_{5\pi}K^+$}}
\newcommand{\etaprkz}{\mbox{$\etapr K^0$}}
\newcommand{\etaprkzd}{\mbox{$\etapr_{\eta\pi\pi} K^0$}}
\newcommand{\etaprkzrg}{\mbox{$\etapr_{\rho\gamma} K^0$}}

\newcommand{\etaprpid}{\mbox{$\etapr_{\eta\pi\pi}\pi^+$}}
\newcommand{\etaprpirg}{\mbox{$\etapr_{\rho\gamma}\pi^+$}}
\newcommand{\etaprpifv}{\mbox{$\etapr_{5\pi}\pi^+$}}

\newcommand{\etaprpizepp}{\mbox{$\etapr_{\eta\pi\pi}\piz$}}
\newcommand{\etaprpizrg}{\mbox{$\etapr_{\rho\gamma}\piz$}}

\newcommand{\etaprkstzd}{\mbox{$\etapr_{\eta\pi\pi} K^{*0}$}}

\newcommand{\etaprkstpd}{\mbox{$\etapr_{\eta\pi\pi} K^{*+}_{K^+\piz}$}}
\newcommand{\etaprkstpkz}{\mbox{$\etapr_{\eta\pi\pi} K^{*+}_{K^0\pi^+}$}}

\newcommand{\etaprrhozd}{\mbox{$\etapr_{\eta\pi\pi}\rho^0$}}

\newcommand{\etaprrhopd}{\mbox{$\etapr_{\eta\pi\pi}\rho^+$}}

\newcommand{\etapretagg}{\mbox{$\etapr_{\eta\pi\pi}\eta_{\gaga}$}}
\newcommand{\etapretathrp}{\mbox{$\etapr_{\eta\pi\pi}\eta_{3\pi}$}}
\newcommand{\etapretarg}{\mbox{$\etapr_{\rho\gamma}\eta_{\gaga}$}}
\newcommand{\etapretargtp}{\mbox{$\etapr_{\rho\gamma}\eta_{3\pi}$}}

\newcommand{\etapretaprd}{\mbox{$\etapr_{\eta\pi\pi}\etapr_{\eta\pi\pi}$}}
\newcommand{\etapretaprrg}{\mbox{$\etapr_{\eta\pi\pi}\etapr_{\rho\gamma}$}}
\newcommand{\Betaprk}{\mbox{$B\ra\etapr K$}}
\newcommand{\Betaprkp}{\mbox{$B^+\ra\etapr K^+$}}
\newcommand{\Betaprkz}{\mbox{$B^0\ra\etapr K^0$}}
\newcommand{\Betaprpi}{\mbox{$B^+\ra\etapr\pi^+$}}
\newcommand{\Betaprpiz}{\mbox{$B^0\ra\etapr\piz$}}
\newcommand{\Betaprkstz}{\mbox{$B^0\ra\etapr K^{*0}$}}
\newcommand{\Betaprkstp}{\mbox{$B^+\ra\etapr K^{*+}$}}
\newcommand{\Betaprrhoz}{\mbox{$B\ra\etapr\rho^0$}}
\newcommand{\Betaprrhop}{\mbox{$B^+\ra\etapr\rho^+$}}
\newcommand{\Betapreta}{\mbox{$B^0\ra\etapr\eta$}}
\newcommand{\Betapretapr}{\mbox{$B\ra\etapr\etapr$}}

\newcommand{\etak}{\mbox{$\eta K^+$}}
\newcommand{\etakgg}{\mbox{$\eta_{\gaga} K^+$}}
\newcommand{\etakthrp}{\mbox{$\eta_{3\pi} K^+$}}

\newcommand{\etapigg}{\mbox{$\eta_{\gaga}\pi^+$}}
\newcommand{\etapithrp}{\mbox{$\eta_{3\pi}\pi^+$}}

\newcommand{\etapizgg}{\mbox{$\eta_{\gaga}\piz$}}
\newcommand{\etapizthrp}{\mbox{$\eta_{3\pi}\piz$}}

\newcommand{\etakzgg}{\mbox{$\eta_{\gaga} K^0$}}
\newcommand{\etakzthrp}{\mbox{$\eta_{3\pi} K^0$}}

\newcommand{\etaetagg}{\mbox{$\eta_{\gaga}\eta_{\gaga}$}}
\newcommand{\etaetathrp}{\mbox{$\eta_{\gaga}\eta_{3\pi}$}}
\newcommand{\etaetasixp}{\mbox{$\eta_{3\pi}\eta_{3\pi}$}}

\newcommand{\etakstzgg}{\mbox{$\eta_{\gaga} K^{*0}$}}
\newcommand{\etakstzthrp}{\mbox{$\eta_{3\pi} K^{*0}$}}

\newcommand{\etakstpgg}{\mbox{$\eta_{\gaga} K^{*+}_{K^+\piz}$}}
\newcommand{\etakstpthrp}{\mbox{$\eta_{3\pi} K^{*+}_{K^+\piz}$}}
\newcommand{\etakstpggkz}{\mbox{$\eta_{\gaga} K^{*+}_{K^0\pi^+}$}}
\newcommand{\etakstpthrpkz}{\mbox{$\eta_{3\pi} K^{*+}_{K^0\pi^+}$}}

\newcommand{\etarhozgg}{\mbox{$\eta_{\gaga} \rho^0$}}
\newcommand{\etarhozthrp}{\mbox{$\eta_{3\pi} \rho^0$}}

\newcommand{\etarhopgg}{\mbox{$\eta_{\gaga} \rho^+$}}
\newcommand{\etarhopthrp}{\mbox{$\eta_{3\pi} \rho^+$}}
\newcommand{\Betak}{\mbox{$B^+\ra\eta K^+$}}
\newcommand{\Betapi}{\mbox{$B^+\ra\eta\pi^+$}}
\newcommand{\Betapiz}{\mbox{$B^0\ra\eta\piz$}}
\newcommand{\Betakz}{\mbox{$B^0\ra\eta K^0$}}
\newcommand{\Betaeta}{\mbox{$B^0\ra\eta\eta$}}
\newcommand{\Betakstz}{\mbox{$B^0\ra\eta K^{*0}$}}
\newcommand{\Betakstp}{\mbox{$B^+\ra\eta K^{*+}$}}
\newcommand{\Betarhoz}{\mbox{$B^0\ra\eta \rho^0$}}
\newcommand{\Betarhop}{\mbox{$B^+\ra\eta \rho^+$}}

\newcommand{\Bomegapi}{\mbox{$B^+\rightarrow\omega\pi^+$}}
\newcommand{\Bomegapiz}{\mbox{$B^0\ra\omega\pi^0$}}
\newcommand{\Bomegak}{\mbox{$B^+\rightarrow\omega K^+$}}
\newcommand{\Bomegakz}{\mbox{$B^0\rightarrow\omega K^0$}}
\newcommand{\Bomegah}{\mbox{$B^+\rightarrow\omega h^+$}}

\newcommand{\Bomegaetapr}{\mbox{$B^0\rightarrow\omega\etapr$}}
\newcommand{\Bomegaeta}{\mbox{$B^0\rightarrow\omega\eta$}}
\newcommand{\Bomegarhoz}{\mbox{$B^0\rightarrow\omega \rho^0$}}
\newcommand{\Bomegarhop}{\mbox{$B^+\rightarrow\omega \rho^+$}}
\newcommand{\Bomegakstz}{\mbox{$B^0\rightarrow\omega K^{*0}$}}
\newcommand{\Bomegakstp}{\mbox{$B^+\rightarrow\omega K^{*+}$}}
\newcommand{\Bomegaomega}{\mbox{$B^0\ra\omega\omega$}}

\newcommand{\omegak}{\mbox{$\omega K^+$}}
\newcommand{\omegakz}{\mbox{$\omega K^0$}}
\newcommand{\omegapi}{\mbox{$\omega\pi^+$}}
\newcommand{\omegapiz}{\mbox{$\omega\pi^0$}}
\newcommand{\omegah}{\mbox{$\omega h^+$}}

\newcommand{\omegaetaprd}{\mbox{$\omega\etapr_{\eta\pi\pi}$}}
\newcommand{\omegaetaprrg}{\mbox{$\omega\etapr_{\rho\gamma}$}}

\newcommand{\omegaetagg}{\mbox{$\omega\eta_{\gaga}$}}
\newcommand{\omegaetathrp}{\mbox{$\omega\eta_{3\pi}$}}

\newcommand{\omegakstzd}{\mbox{$\omega K^{*0}_{K^+\pi^-}$}}

\newcommand{\omegakstpd}{\mbox{$\omega K^{*+}_{K^+\piz}$}}
\newcommand{\omegakstpkz}{\mbox{$\omega K^{*+}_{K^0\pi^+}$}}
\newcommand{\omegarhoz}{\mbox{$\omega \rho^0$}}
\newcommand{\omegarhop}{\mbox{$\omega \rho^+$}}
\newcommand{\omegaomega}{\mbox{$\omega\omega$}}

\newcommand{\Bphik}{\mbox{$B^+\ra\phi K^+$}}
\newcommand{\Bphikz}{\mbox{$B^0\ra\phi K^0$}}
\newcommand{\Bphipi}{\mbox{$B^+\ra\phi\pi^+$}}
\newcommand{\Bphipiz}{\mbox{$B^0\ra\phi\pi^0$}}
\newcommand{\Bphietapr}{\mbox{$B^0\ra\phi\etapr$}}
\newcommand{\Bphieta}{\mbox{$B^0\ra\phi\eta$}}
\newcommand{\Bphikstz}{\mbox{$B^0\ra\phi K^{*0}$}}
\newcommand{\Bphikstp}{\mbox{$B^+\ra\phi K^{*+}$}}
\newcommand{\Bphikstpd}{\mbox{$B^+\ra\phi K^{*+}(K^{*+}\ra K^+\piz$}}
\newcommand{\Bphikstpkz}{\mbox{$B^+\ra\phi K^{*+} (K^{*+}\ra K^0\pi^+$}}
\newcommand{\Bphikst}{\mbox{$B\ra\phi K^*$}}
\newcommand{\Bphirhoz}{\mbox{$B^0\ra\phi \rho^0$}}
\newcommand{\Bphirhop}{\mbox{$B^+\ra\phi \rho^+$}}
\newcommand{\Bphiomega}{\mbox{$B^0\ra\phi\omega$}}
\newcommand{\Bphiphi}{\mbox{$B^0\ra\phi\phi$}}

\newcommand{\phik}{\mbox{$\phi K^+$}}
\newcommand{\phikz}{\mbox{$\phi K^0$}}
\newcommand{\phipi}{\mbox{$\phi\pi^+$}}
\newcommand{\phipiz}{\mbox{$\phi\pi^0$}}

\newcommand{\phietaprd}{\mbox{$\phi\etapr_{\eta\pi\pi}$}}
\newcommand{\phietaprrg}{\mbox{$\phi\etapr_{\rho\gamma}$}}

\newcommand{\phietagg}{\mbox{$\phi\eta_{\gaga}$}}
\newcommand{\phietathrp}{\mbox{$\phi\eta_{3\pi}$}}

\newcommand{\phikstzd}{\mbox{$\phi K^{*0}_{K^+\pi^-}$}}
\newcommand{\phikstzkz}{\mbox{$\phi K^{*0}_{K^0\piz}$}}

\newcommand{\phikstpd}{\mbox{$\phi K^{*+}_{K^+\piz}$}}
\newcommand{\phikstpkz}{\mbox{$\phi K^{*+}_{K^0\pi^+}$}}

\newcommand{\phirhoz}{\mbox{$\phi \rho^0$}}
\newcommand{\phirhop}{\mbox{$\phi \rho^+$}}
\newcommand{\phiomega}{\mbox{$\phi\omega$}}
\newcommand{\phiphi}{\mbox{$\phi\phi$}}

\newcommand{\etaprinc}{\mbox{$B\ra\etapr X_S$}}
\def\babar{{\sl B}$\scriptstyle\sl A${\sl B}$\scriptstyle\sl AR$}

\def\displaytwo#1#2#3#4#5{ 
   \begin{figure}[#1]
   \centerline{\psfig{figure=#2.ps,height=3.6in,clip=}} 
   \Caption{#4}\vskip 1.0truecm
   \label{#2}
   \centerline{\psfig{figure=#3.ps,height=3.6in,clip=}}
   \Caption{#5}
   \label{#3}
   \end{figure}}

\def\Caption#1{\renewcommand{\baselinestretch}{1.0} \caption{#1} 
                         \renewcommand{\baselinestretch}{1.667} }

\begin{document}

\preprint{\tighten\vbox{\hbox{\hfil COLO--HEP--395}
                        \hbox{\hfil \date{\today}}
\vskip 3.0truecm}}

\title{
First observation of five charmless hadronic $B$ decays}

\tighten
\author{\tighten J.\ G.\ Smith
 \address{Department of Physics, University of Colorado\\
  Boulder, Colorado 80309-0390}%
 \thanks{Invited talk presented at the {\sl Seventh International Symposium
On Heavy Flavor Physics}, Santa Barbara, CA, July 7-11, 1997.}
}
\maketitle
\tighten

\begin{abstract}
There has been much progress in measurements of charmless hadronic $B$
decays during 1997.  Building on the previous indications from CLEO
and LEP, CLEO now has clear signals in five exclusive final states:
$K^+\pi^-$, $K^0\pi^+$, $\etapr K^+$, $\etapr K^0$, and $\omega K^+$.
The branching fractions for the $\etapr K$ modes are several
times larger than the others.  A similar strikingly large signal has
been seen in the inclusive decay, $B\ra\etapr X_S$.  All of these
signals would appear to be dominated by hadronic penguin processes.
\end{abstract}
\newpage
  
\section{Introduction}

\begin{figure}[htbp]
\centering
\leavevmode
\epsfxsize=6.5in
\epsffile{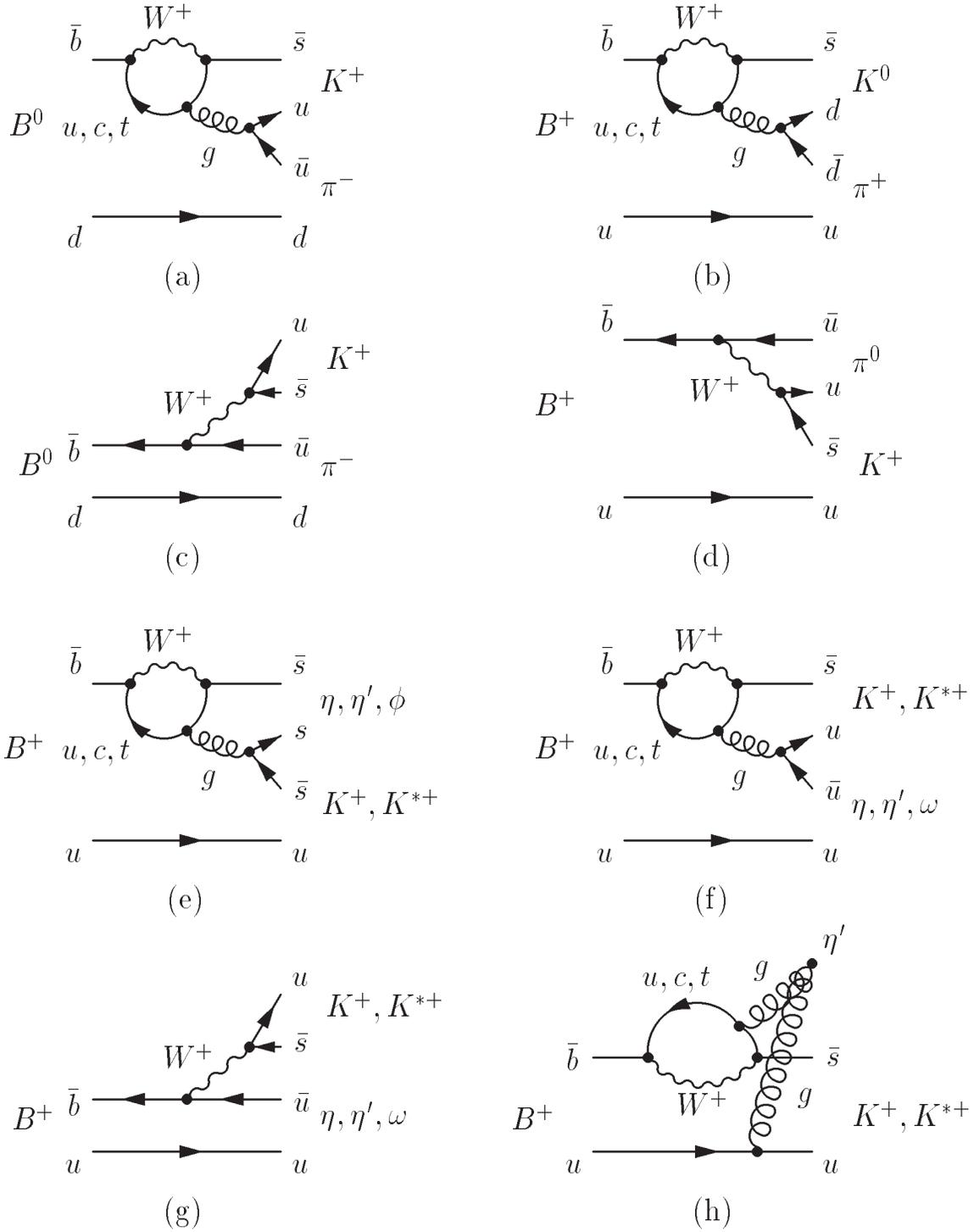}\bigskip
\caption{Feynman diagrams for some of the penguin and spectator processes 
which are expected to be dominant for the modes described in this paper.}
\label{feynfig}
\end{figure}

Charmless hadronic $B$ decays are expected to proceed primarily through 
$b\ra s$ loop (``penguin") diagrams and $b\ra u$ spectator diagrams.  In
Fig. \ref{feynfig} we show four such diagrams for three of the $K\pi$
modes discussed in this paper.  We also show the diagrams expected to
dominate the final states with an isoscalar meson and a $K$ or $K^*$
meson.  Interchange of $d$ and $u$ spectator quarks will generally
provide the diagrams for both $B^+$ and $B^0$ decays.  Diagrams
\ref{feynfig}c, \ref{feynfig}d, and \ref{feynfig}g are Cabibbo
suppressed.  The un-suppressed versions of these diagrams and the CKM
\cite{ckm} suppressed versions of the penguin diagrams lead to final states
such as $\pi\pi$, $\etapr\pi$, $\eta\rho$, and $\omega\pi$.

These decays have received a great deal of attention because interference
among penguin and spectator diagrams leading to the same final state can
produce (direct) $CP$ violation, and, for the $B^0$ system, interference
between final states reached directly or via $B$-$\bar B$ mixing, can 
generate (indirect) $CP$ violation \cite{buras}. Detectors such as \babar,
Belle or others at hadron colliders will attempt to measure 
an oscillation in the time evolution of certain $B^0$ decays, which is
sensitive to the value of some of the CKM angles.
Other approaches have been suggested \cite{deshhe}
notably the possibility of using
``quadrangle" relations \cite{rosner} among amplitudes of four related decays 
to determine CKM angles.

In subsequent sections, we will review the experimental situation prior
to 1997 and then report the results of several new analyses from CLEO
which have found the first unambiguous evidence for individual charmless
hadronic $B$ decays.  We conclude with interpretations of these results.

\section{Previously published results}

Until this year, there have been relatively few indications of charmless
hadronic $B$ decays.  CLEO first published evidence for the modes
\Bkpi\ and \Bpipi\ \cite{CLEObkpi}, but due to lack of statistics, was
unable to claim a signal for either mode separately.  Subsequently CLEO
updated these results, still without an observation of either mode
individually, and provided limits for many other related modes \cite{bigrare}.

\begin{figure}[htbp]
\label{alephfig}
\begin{center}
\epsfig{figure=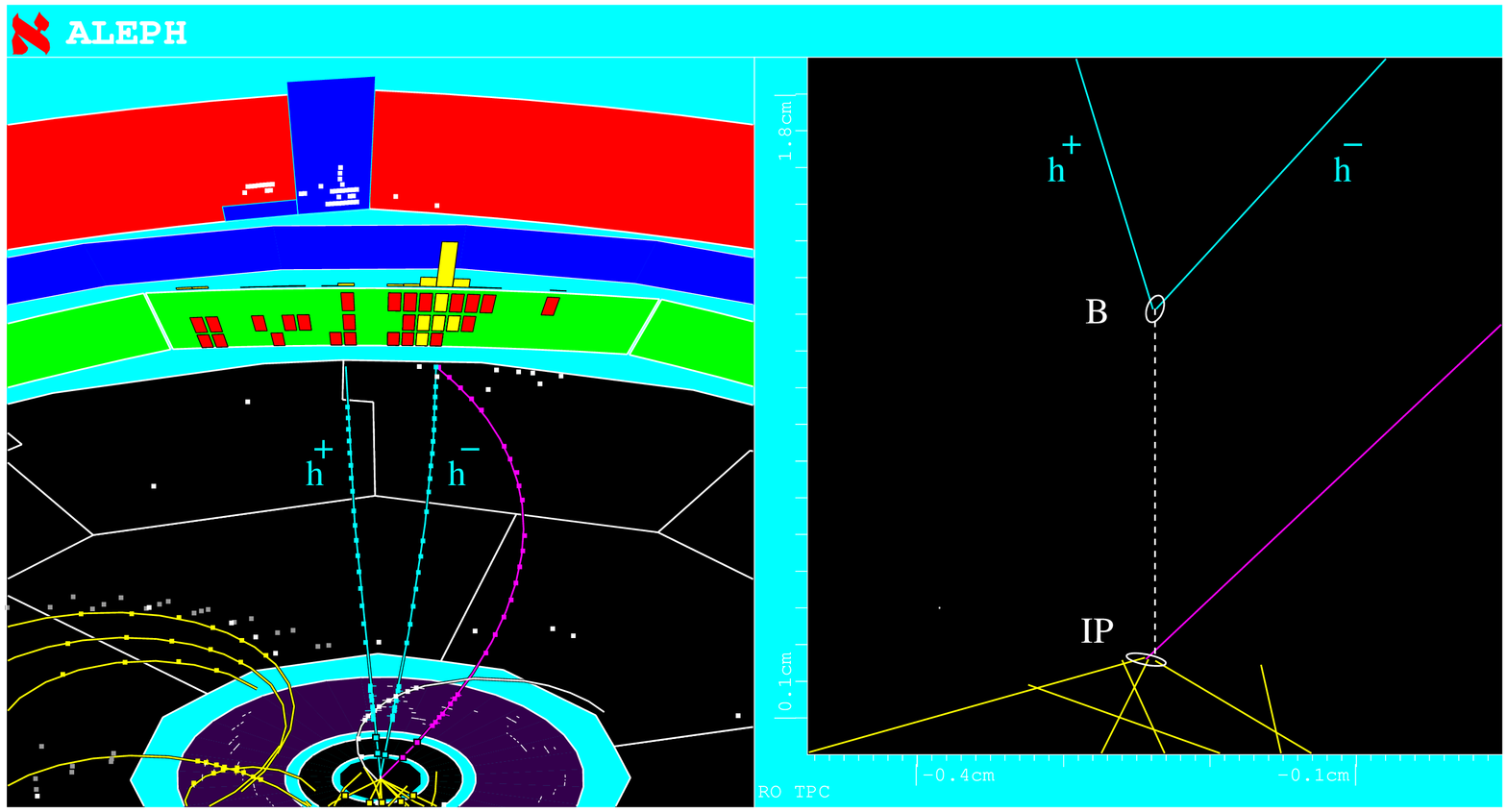,height=3.5in,angle=270}\bigskip
\caption{Event from the ALEPH experiment showing a \Bkpi\ decay
candidate as reconstructed with use of the ALEPH silicon vertex detector.}
\end{center}
\bigskip
\begin{center}
\epsfig{figure=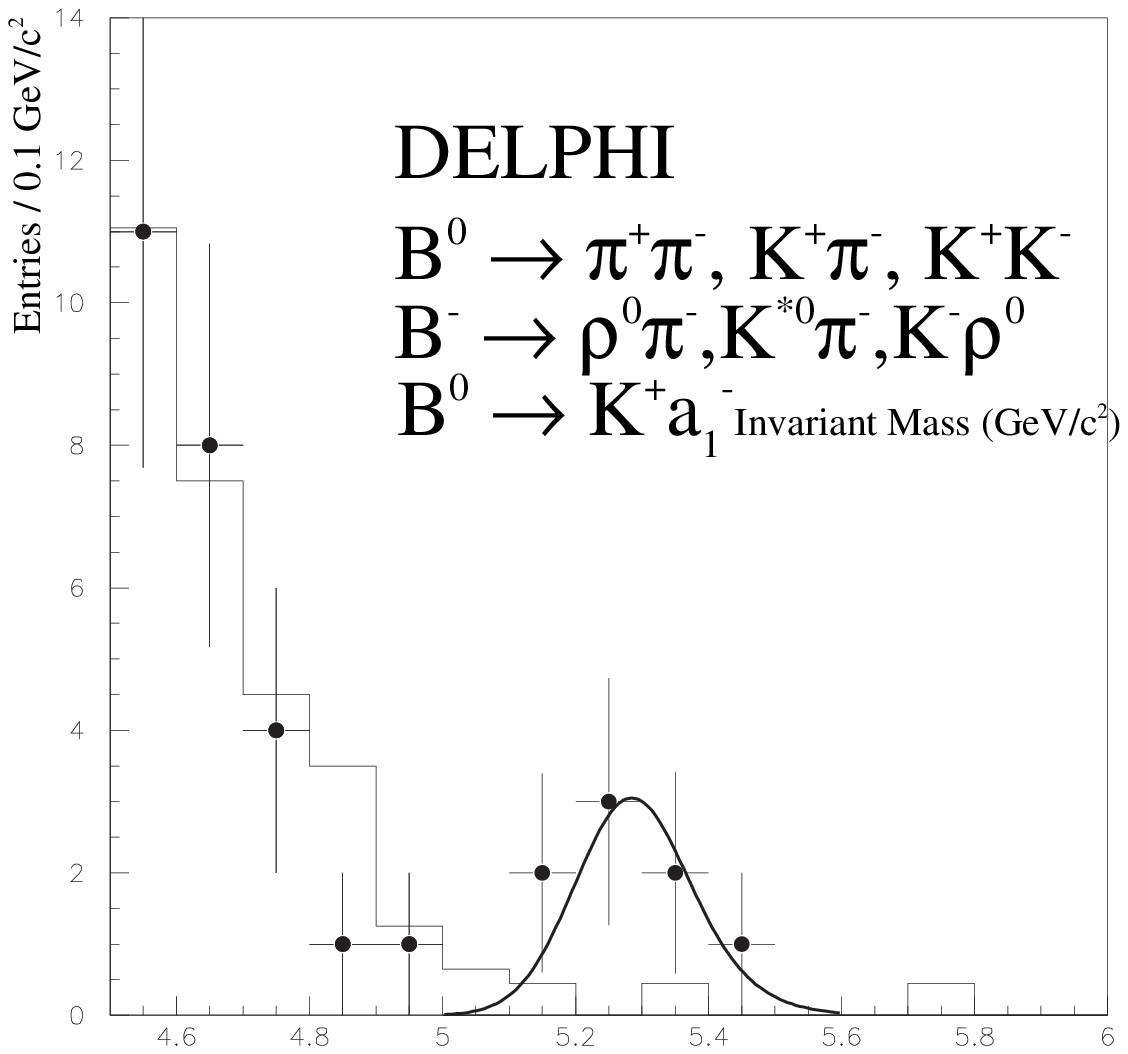,height=3.5in}\bigskip
\caption{Invariant mass distribution obtained by DELPHI for 
ten charmless hadronic $B$ decay candidates, background events at lower mass, 
and the Monte Carlo expectation for the background.}
\label{delphifig}
\end{center}
\end{figure}

At LEP $B$ mesons are produced with high momentum so they travel $\sim$1
mm before decaying.  Several LEP experiments have used the excellent
vertex resolution provided by their silicon vertex detectors to obtain
virtually background-free evidence for charmless hadronic $B$ decays.
Examples of such events are shown in Fig. \ref{alephfig} for the ALEPH
experiment \cite{aleph}. The experimental difficulty is that the LEP experiments
also cannot separate the decays \Bkpi\ and \Bpipi, or these from
$B_S\ra K^+K^-$.  Fig. \ref{delphifig} shows
the mass distribution for ten candidate charmless hadronic $B$ decays
from the DELPHI experiment \cite{delphi}.

\section{Results for exclusive decays from CLEO}

The results in this section are based on data collected with the CLEO II
detector \cite{CLEOdet} at the Cornell Electron Storage Ring (CESR). The data
sample corresponds to an integrated luminosity of 3.11~fb$^{-1}$ taken on the
$\Upsilon$(4S) resonance and 1.61~fb$^{-1}$ taken slightly below. The
on-resonance data sample contains $3.3 \times 10^6$ $B\overline{B}$ pairs.
Resonance states are reconstructed from charged tracks and photons
with the decay channels: \etaprepp, \etaprrg, $K^0$ via $K_S\ra\pi^+ \pi^-$,
$\rho^0\ra\pi^+\pi^-$, $\pi^0 \ra \gamma\gamma$, \etagamgam, \etathreepi,
\omegappp, $\phi\ra K^+K^-$, $K^{*0}\ra K^+\pi^-$, $K^{*0}\ra K^0\pi^0$,
$K^{*+}\ra K^+\pi^0$, and $K^{*+} \ra K^0\pi^+$.  Charge conjugate
decays are implied throughout this paper.

Candidate charged tracks are required to pass quality cuts and have
specific ionization ($dE/dx$) consistent with that of a
pion and kaon.  Such tracks must not be electrons
or muons, identified by calorimetry and depth of penetration of an iron
muon stack, respectively.  \kshpp\ candidates are accepted
only if they are displaced from the primary interaction point
by at least $3$~mm.
Photon candidates are isolated calorimeter showers with a measured 
energy of at least 30 (50) MeV in the central (end cap) region of the
calorimeter.  The momentum of charged tracks and photon pairs is
required to be greater than 100~MeV/c to reduce combinatoric background.
Photon pairs and vees are fit kinematically to the appropriate combined
mass hypothesis to obtain the meson momentum vectors.  Resolutions on
the reconstructed masses prior to the constraint are about 5-10 MeV
for $\piz\ra\gaga$, 12 MeV for \etagg, and 3 MeV for \kshpp.

The primary means of identification of $B$ meson candidates is through their
measured mass and energy. The resolution for $\Delta E
\equiv E_1 + E_2 - E_b$ ($E_1$ and $E_2$ are the energy of the two
daughter particles of the $B$ and $E_b$ is the beam energy) is typically
25-50 MeV.  The resolution for $M \equiv \sqrt{E_b^2 - {\bf p}_B^2}$ 
(${\bf p}_B$ is the reconstructed $B$ momentum) is
2.5-3.0 MeV, dominated by the uncertainty in ${\bf p}_B$.  
Signals are identified with the use of resonance masses and, in the case of
vector-pseudoscalar decays and the \etaprrg\ channel, a variable \hel\ 
sensitive to the helicity distribution of the decay.
For modes in which one daughter is a single charged track, or is a 
resonance pairing a charged track with a $\pi^0$, the $dE/dx$ variables
$S_K$ and $S_\pi$ are used.  The latter are defined as the
deviations from nominal energy loss for the indicated particle
hypotheses measured in standard deviations.
Studies of $D^{*+}$ tagged
$D^0\ra K^-\pi^+$ decays find a $K$-$\pi$ separation of
about 1.7 standard deviations near 2.5 GeV/c.

The large background from continuum quark production ($e^+e^-\ra
q\bar q$) can be reduced with the use of event shape cuts.  One such cut
involves the quantity $\theta_{BB}$, the angle between the thrust axis
of the candidate $B$ and that of the rest of the event (sphericity is 
used instead of thrust for the $K\pi$ and $\pi\pi$ analyses).  Since $B$ 
mesons are produced nearly at rest, there is little correlation between the 
two thrust axes, while candidates extracted from continuum $q\bar q$ events 
tend to be strongly correlated by the jet-like nature of the events.
This difference is
exploited by requiring $\left|\cos\theta_{BB}\right|<0.9$.
A multivariate discriminant \xf\ is also employed, with the
primary inputs being the energy deposition in nine cones concentric 
with the thrust or sphericity axis of the candidates tracks.  Monte Carlo
studies indicate that backgrounds from other $B$ decay modes are small
and they are not considered further.

In order to extract event yields, an unbinned extended-maximum-likelihood (ML) 
fit \cite{mlfit} is
performed, which includes sidebands about the expected mass and energy peaks,
of a superposition of expected signal and background distributions:
\begin{eqnarray*}
 {\cal L} & = & e^{-(N_S+N_B)} \prod_{i=1}^N \{N_{S} {\cal P}_{S_i}
   (f_1,...,f_m;x_1,...x_p) + 
   N_B {\cal P}_{B_i}(g_1,...,g_m;x_1,...x_p)\} ,
\end{eqnarray*}
where ${\cal P}_{S_i}$ and ${\cal P}_{B_i}$ are the probabilities for event
$i$ to be signal and continuum background, respectively. The probabilities are 
a function of the values of the variables $x$ used in the fit for each event, 
and of the parameters $f$ and $g$ used to describe the signal and background
shapes for each variable. The variables used are $\Delta E$, $M$, ${\cal F}$,
and, where applicable, resonance masses, ${\cal H}$, $S_K$, and $S_\pi$.
$N_{S}$ and $N_B$, the free parameters of the fit, are the (positive-definite)
number of signal and continuum background events in the fitted data sample,
respectively.  Sample sizes for these fits range from $\sim 30$ to about
ten thousand events.

The signal probability distribution functions (PDFs) 
$\cal P_S$ and $\cal P_B$ are constructed as products of
functions of the observables ${\bf x}_i$; 
they are determined from fits to Monte Carlo events that simulate the
response of the CLEO detector to each decay mode investigated. The
GEANT \cite{geant} based simulation is tuned to reproduce detector resolution
and efficiencies for a variety of benchmark processes. The parameters of the
background PDFs are determined with similar fits to a sideband region of data
defined by $|\Delta E| < 0.2$~GeV and $5.2 < M < 5.27$~GeV/c$^2$. The
signal shapes used are Gaussian, double Gaussian, and Breit-Wigner as
appropriate for $\Delta E$ and mass peaks. For background, resonance masses
are fit to the sum of a smooth polynomial and the signal shape, to account
for the component of real resonance as well as the combinatoric background. 
Shapes used for $\Delta E$ and $M$ background are, respectively, a first-degree
polynomial and the empirical shape \cite{argus} $f(z) \propto
z\sqrt{1-z^2}\exp{(-\xi(1-z^2))}$, where $z\equiv M/E_b$ and $\xi$ is a
parameter to be fit. Finally, for ${\cal F}$, $S_K$, and
$S_{\pi}$, bifurcated Gaussians are used for both signal and background.

\begin{table}[htbp]
\begin{center}
\caption{Experimental results and theoretical predictions. 
Columns list the event yield from the fit, statistical significance,
reconstruction efficiency $\epsilon$ (including the branching fraction
for the $K^0$ via $K_S\ra\pi^+ \pi^-$ chain), and the 
resulting $B$ decay branching fraction ${\cal B}$.}
\vspace{0.2cm}
\begin{tabular}{lcccccl}
\dbline
Final&ML fit&&&&Theory&\\
state& events & Signif. & $\epsilon$(\%) &\calB($10^{-5})$
& \calB($10^{-5})$&References\\
\hline
$\pi^+\pi^-$ &$9.9^{+6.0}_{-5.1}$  & 2.2$\sigma$        &$44\pm3$
& $<~1.5$  & 0.8--1.8 & \ref{chau},\ref{dean},\ref{kp},\ref{du}    \\
$\pi^+\pi^0$ & $11.3^{+6.3}_{-5.2}$ & 2.8$\sigma$        &$37\pm3$
& $<~2.0$  & 0.6--2.0 & \ref{chau},\ref{dean},\ref{kp},\ref{du}      \\
$\pi^0\pi^0$ & $2.7^{+2.7}_{-1.7}$ & 2.4$\sigma$        &$29\pm3$
& $<0.93$ & 0.02--0.06& \ref{chau},\ref{dean},\ref{kp},\ref{du}      \\
\hline
$K^+\pi^-$   &$21.6^{+6.8}_{-6.0}$ & 5.6$\sigma$         &$44\pm3$
&$1.5^{+0.5}_{-0.4}\pm0.1$ & 0.7--2.4 & \ref{desh}-\ref{dean},\ref{kp},\ref{du} \\
$K^+\pi^0$   &$8.7^{+5.3}_{-4.2}$  & 2.7$\sigma$        &$37\pm3$
& $<~1.6$  & 0.3--1.3 & \ref{desh}-\ref{dean},\ref{kp},\ref{du} \\
$K^0\pi^+$   &$9.2^{+4.3}_{-3.8}$  & 3.2$\sigma$       &$12\pm1$
&$2.3^{+1.1}_{-1.0}\pm0.4$ &  0.5--1.3    & \ref{desh}-\ref{sw},\ref{fl},\ref{kp},\ref{du} \\
$K^0\pi^0$   & $4.1^{+3.1}_{-2.4}$ & 2.2$\sigma$        &$8\pm1$ & $<~4.1$  
& 0.2--0.8 & \ref{desh},\ref{chau},\ref{dean},\ref{kp},\ref{du}\\
\hline
$K^+K^-$     &$0.0^{+1.3}_{-0.0}$   & 0.0$\sigma$       &$44\pm3$
& $<~0.43$ &       --      &        \\
$K^+\bar{K}^0$   & $0.6^{+3.8}_{-0.6}$  & 0.2$\sigma$       &$12\pm1$
& $<~2.1$  & 0.06--0.24 & \ref{chau},\ref{sw},\ref{fl},\ref{kp},\ref{du} \\
$K^0\bar{K}^0$   & 0   & --      &$5\pm 1$
& $<~1.7$  & 0.06--0.13 & \ref{chau},\ref{fl},\ref{kp},\ref{du} \\
\hline
$h^+\pi^0$   &$20.0^{+6.8}_{-5.9}$ & 5.5$\sigma$         &$37\pm3$
&$1.6^{+0.6}_{-0.5}\pm0.4$  & --   &  \\
\dbline
\end {tabular}
\label{kpitab}
\end{center}
\end {table}

\begin{figure}[hbtp]
\begin{center}
\epsfig{figure=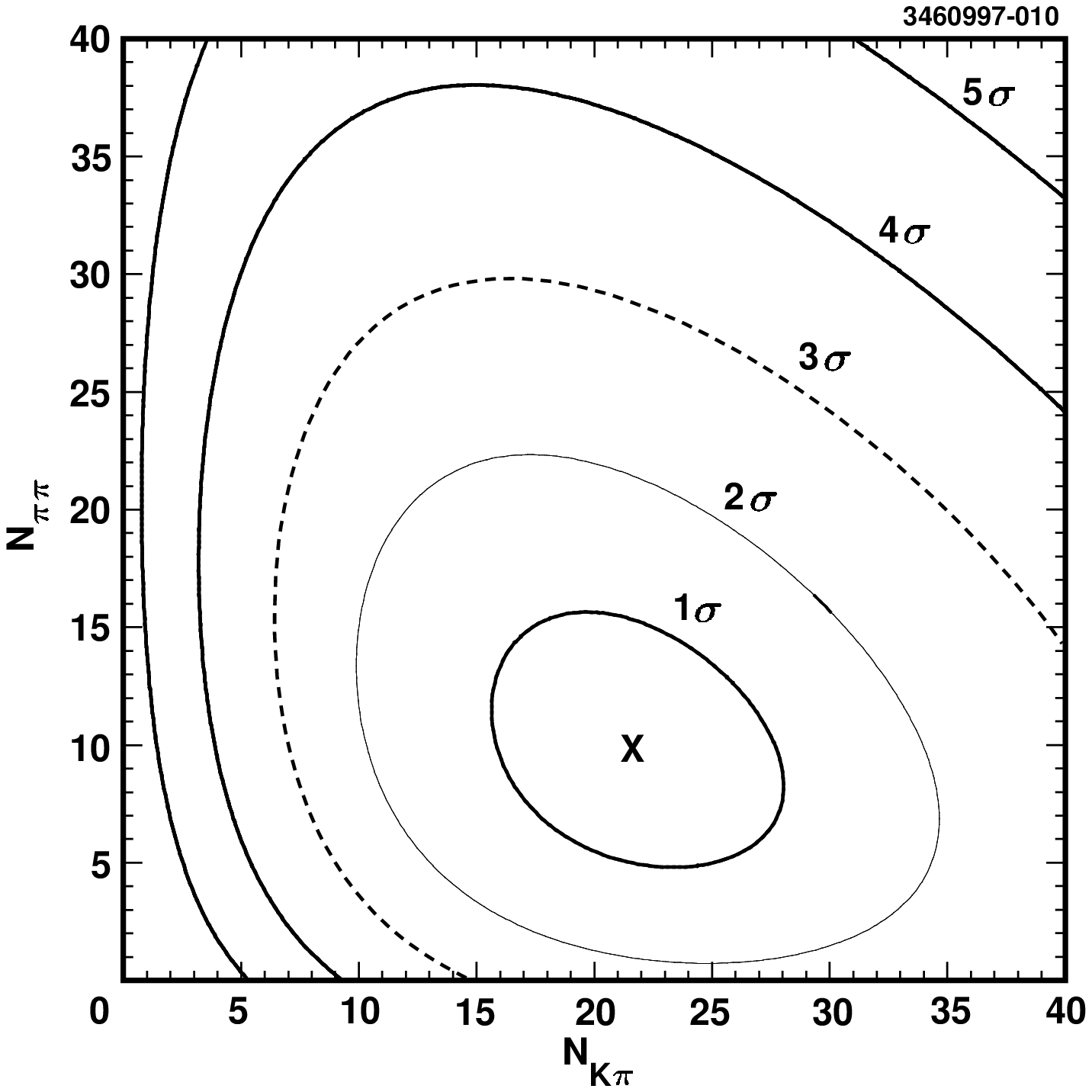,height=2.1in}
\epsfig{figure=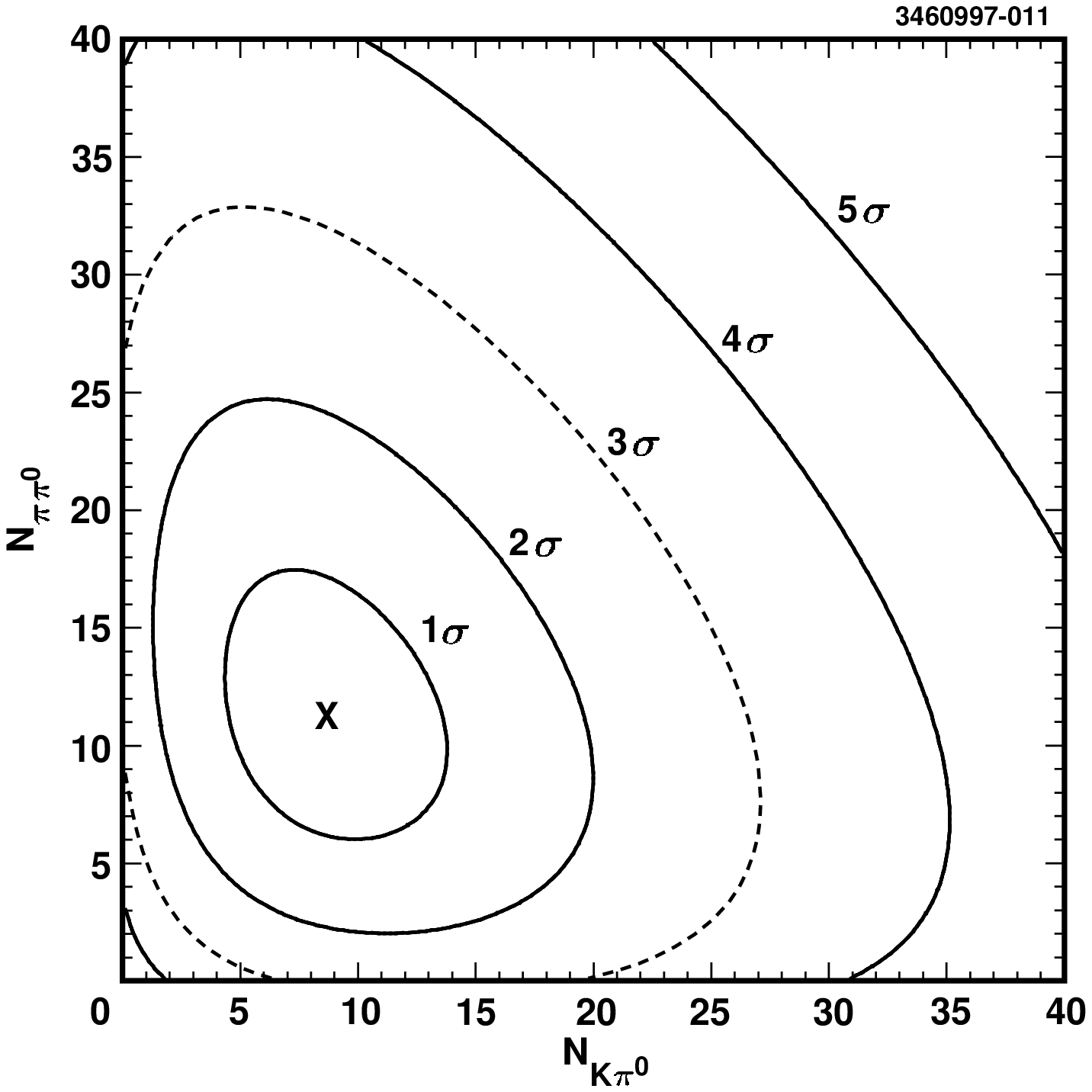,height=2.1in}
\epsfig{figure=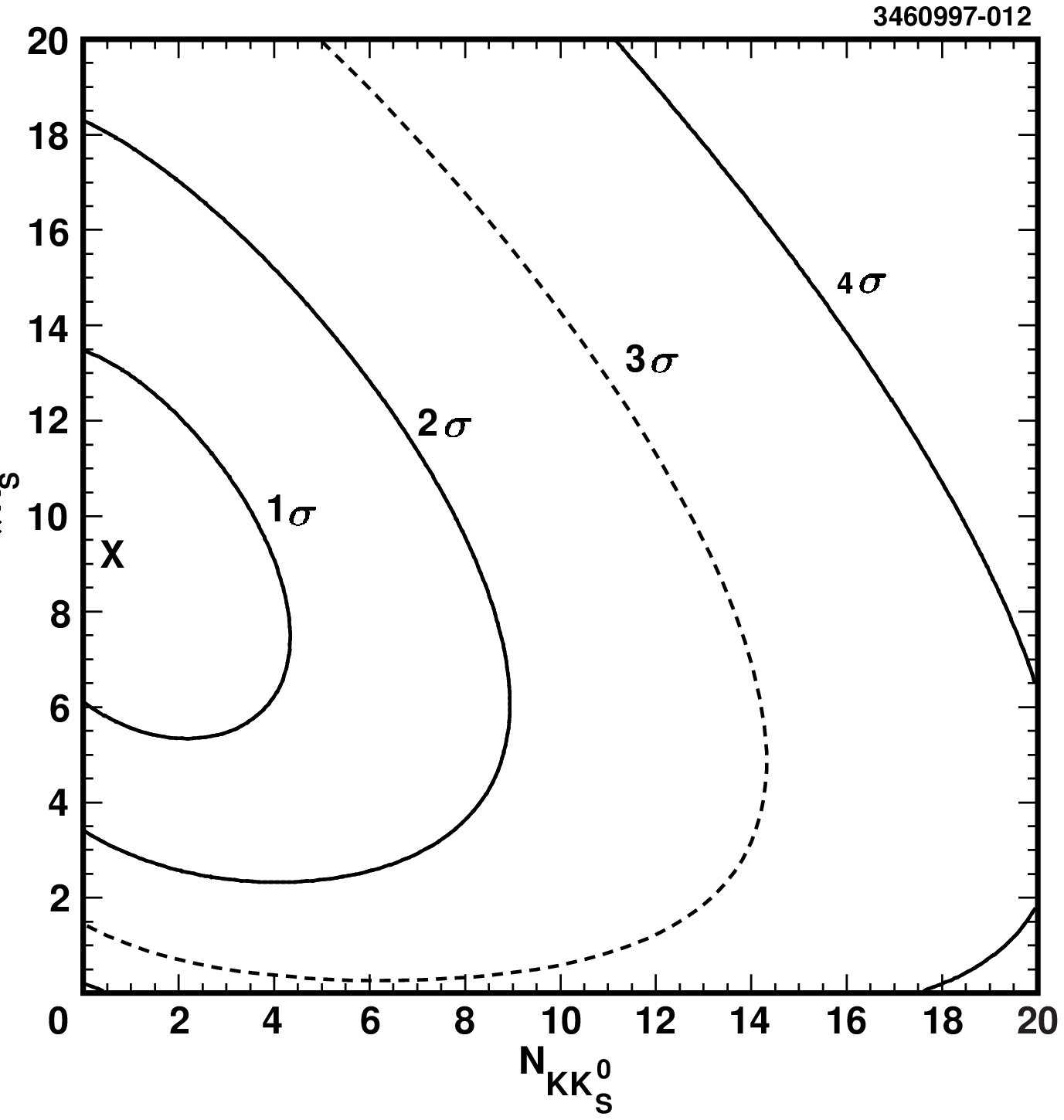,height=2.1in}
\end{center}
\caption{Likelihood contours for ML fits to
(a) $N_{K^{\pm}\pi^{\mp}}$ and $N_{\pi^+\pi^-}$ for $B^0\ra K^+\pi^-$
and $B^0\ra \pi^+\pi^-$; (b) $N_{K\pi^0}$ and $N_{\pi\pi^0}$ for
$B^+\ra K^+\pi^0$ and $B^+\ra \pi^+\pi^0$; (c) $N_{K^0_SK}$ and $N_{K^0_S\pi}$ 
for $B^+\ra \bar{K}^0K^+$ and $B^+\ra K^0\pi^+$.}
\label{contourkpi}
\begin{center}
\epsfig{figure=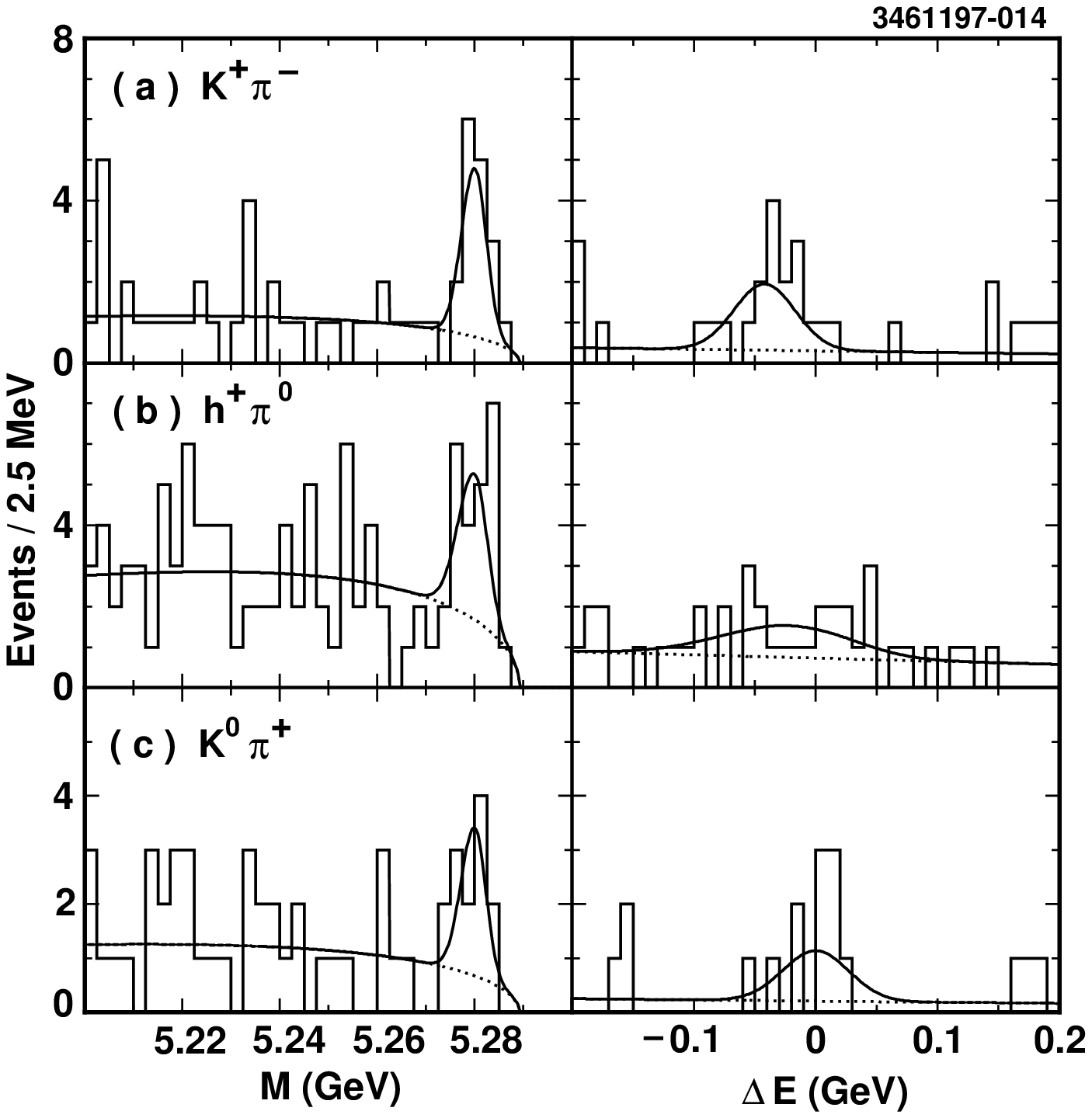,height=4.5in}
\end{center}
\caption{$M$ and $\Delta E$ plots for (a) $B^0\ra K^+\pi^-$,
(b) $B^+\ra h^+\pi^0$, and (c) $B^+\ra K^0\pi^+$.
The scaled projection of the total likelihood fit (solid curve)
and the continuum background component (dotted curve) are overlaid.}
\label{kpiproj}
\end{figure}

The results of these fits are given in a Tables \ref{kpitab}-\ref{omcombtab}, 
each with the signal event yield, the efficiency including secondary
branching fractions, and the branching fraction for
each mode, given as a central value with statistical and systematic error or
as a 90\% confidence level upper limit.  Table \ref{kpitab} for the
$K\pi$ and $\pi\pi$ final states also contains the statistical significance 
for the fit to each mode.  Systematic errors in yield and efficiency are 
estimated by variation of the fit parameters and estimation of uncertainties in
reconstruction efficiencies and selection requirements.  Branching fraction 
upper limits are obtained by increasing the yield and reducing the
efficiency by their systematic errors.  In Fig.~\ref{contourkpi}, we show
the ML contour for the three cases in Table \ref{kpitab} with
significance greater than three standard deviations.  While the
significance for the $K^+\piz$ and $\pi^+\piz$ final states are both
(barely) below $3\sigma$, there is strong evidence for $h\piz$, where
$h$ indicates a charged $K$ or $\pi$.  This is very similar to the
case of the $h^+h^-$ final state several years ago \cite{CLEObkpi}.  In
Fig.~\ref{kpiproj}, we show projections of the fit onto the $M$ and
\DE\ axes; cuts have been made on other ML variables in order to better
reflect the background near the signal region.  Further details can be
found in reference \ref{kpipub}.

\begin{table}[htbp]
\caption{Measurement results for $\etapr$ decay modes.  Columns list the final 
states (with secondary decay modes as subscripts), event yield from the fit,
reconstruction efficiency $\epsilon$, total efficiency with secondary
branching fractions ${\cal B}_s$, and the resulting $B$ decay branching
fraction ${\cal B}$.  }
\vspace{0.2cm}
\begin{center}
\begin{tabular}{lcrrc}
\dbline
Final state&Fit events&$\epsilon$(\%)&$\epsilon\calB_s$(\%)&\calB($10^{-5})$\cr
\sgline
\etaprkpd     &$11.2^{+4.1}_{-3.4}$ &30&5.1&$6.7^{+2.5}_{-2.1}\pm0.8$\cr
\etaprkprg    &$19.6^{+6.6}_{-5.7}$ &28&8.4&$7.0^{+2.4}_{-2.1}\pm0.9$\cr
\etaprkpfv    & $2.3^{+2.2}_{-1.5}$ &17&1.7&$4.2^{+4.0}_{-2.7}\pm1.4$\cr
\etaprkzd     & $1.4^{+1.7}_{-1.0}$ &23&1.4&$3.1^{+3.7}_{-2.1}\pm0.6$\cr
\etaprkzrg    & $5.7^{+3.7}_{-2.8}$ &27&2.8&$6.2^{+4.0}_{-3.0}\pm1.2$\cr
\etaprpid     & $1.4^{+2.2}_{-1.4}$ &30&      5.2      & $<3.7$\cr
\etaprpirg    & $4.0^{+4.6}_{-3.3}$ &29&      8.8      & $<4.5$\cr
\etaprpifv    & $0.5^{+1.9}_{-0.5}$ &18&      1.8      & $<10.7$\cr
\etaprpizepp  & $0.0^{+0.5}_{-0.0}$ &25&      4.3      & $<1.8$\cr
\etaprpizrg   & $0.0^{+2.0}_{-0.0}$ &29&      8.7      & $<2.2$\cr
\etapretaprd  & $0.0^{+0.5}_{-0.0}$ &19&      0.6      &$<15.2$\cr
\etapretaprrg & $0.0^{+0.8}_{-0.0}$ &19&      1.7      & $<6.4$\cr
\etapretagg   & $0.0^{+0.5}_{-0.0}$ &26&      1.8      & $<4.6$\cr
\etapretathrp & $0.0^{+0.5}_{-0.0}$ &17&      0.7      &$<12.5$\cr
\etapretarg   & $5.6^{+4.6}_{-3.6}$ &28&      3.3      &$<13.0$\cr
\etapretargtp & $0.0^{+0.6}_{-0.0}$ &16&      1.1      & $<9.3$\cr
\etaprkstpd   & $0.0^{+1.0}_{-0.0}$ &13&      0.7      &$<18.$\cr
\etaprkstpkz  & $0.0^{+1.6}_{-0.0}$ &15&      0.6      &$<24.$\cr
\etaprkstzd   & $0.0^{+0.7}_{-0.0}$ &22&      2.5      & $<3.9$\cr
\etaprrhopd   & $0.0^{+0.7}_{-0.0}$ &12&      2.0      & $<5.7$\cr
\etaprrhozd   & $0.0^{+0.5}_{-0.0}$ &22&      3.8      & $<2.3$\cr
\dbline
\end{tabular}
\end{center}
\label{individtaba}
\end{table}

\begin{figure}[htbp]
\psfiletwoBB{60 150 530 610}{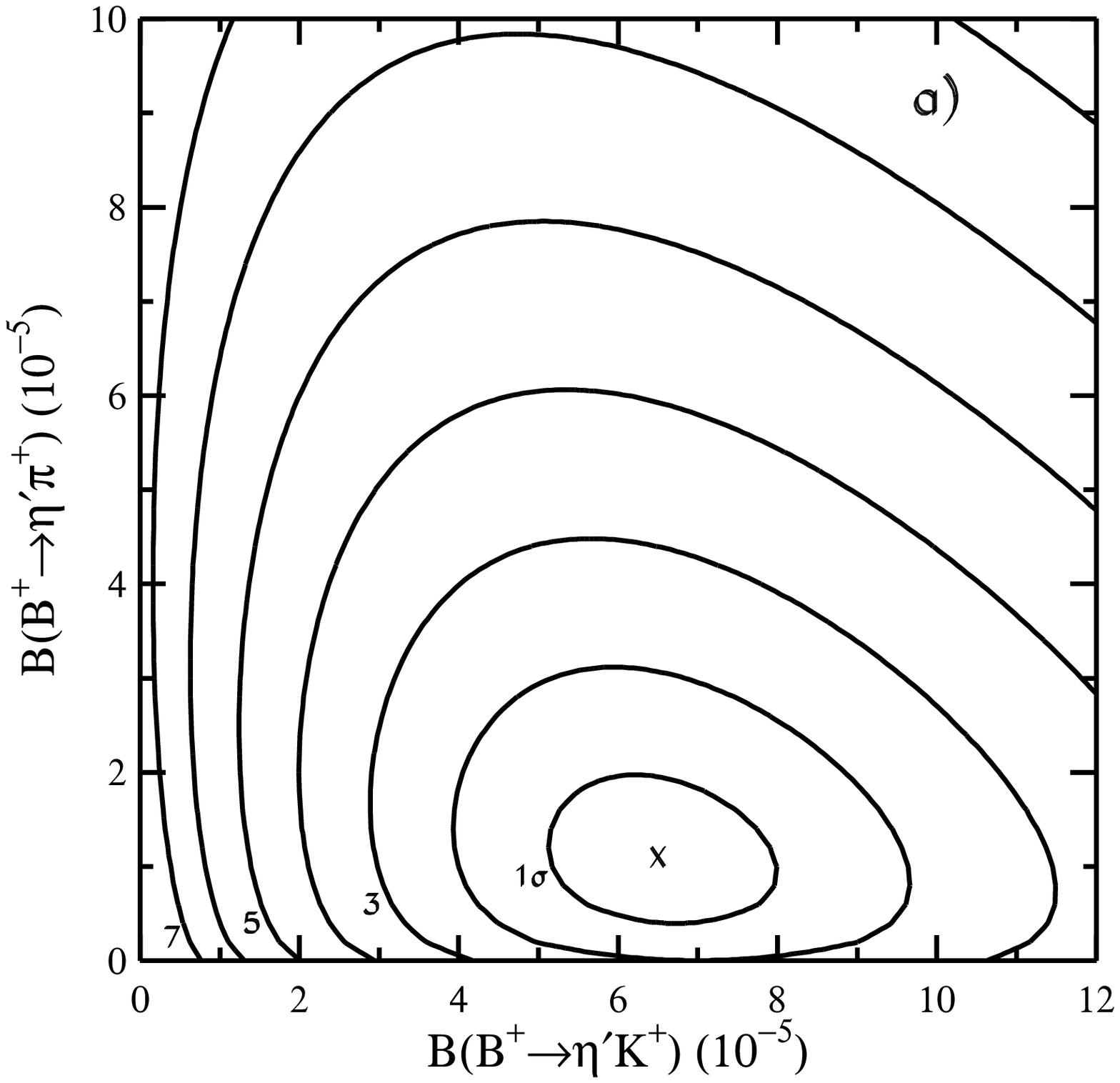}%
{60 150 530 610}{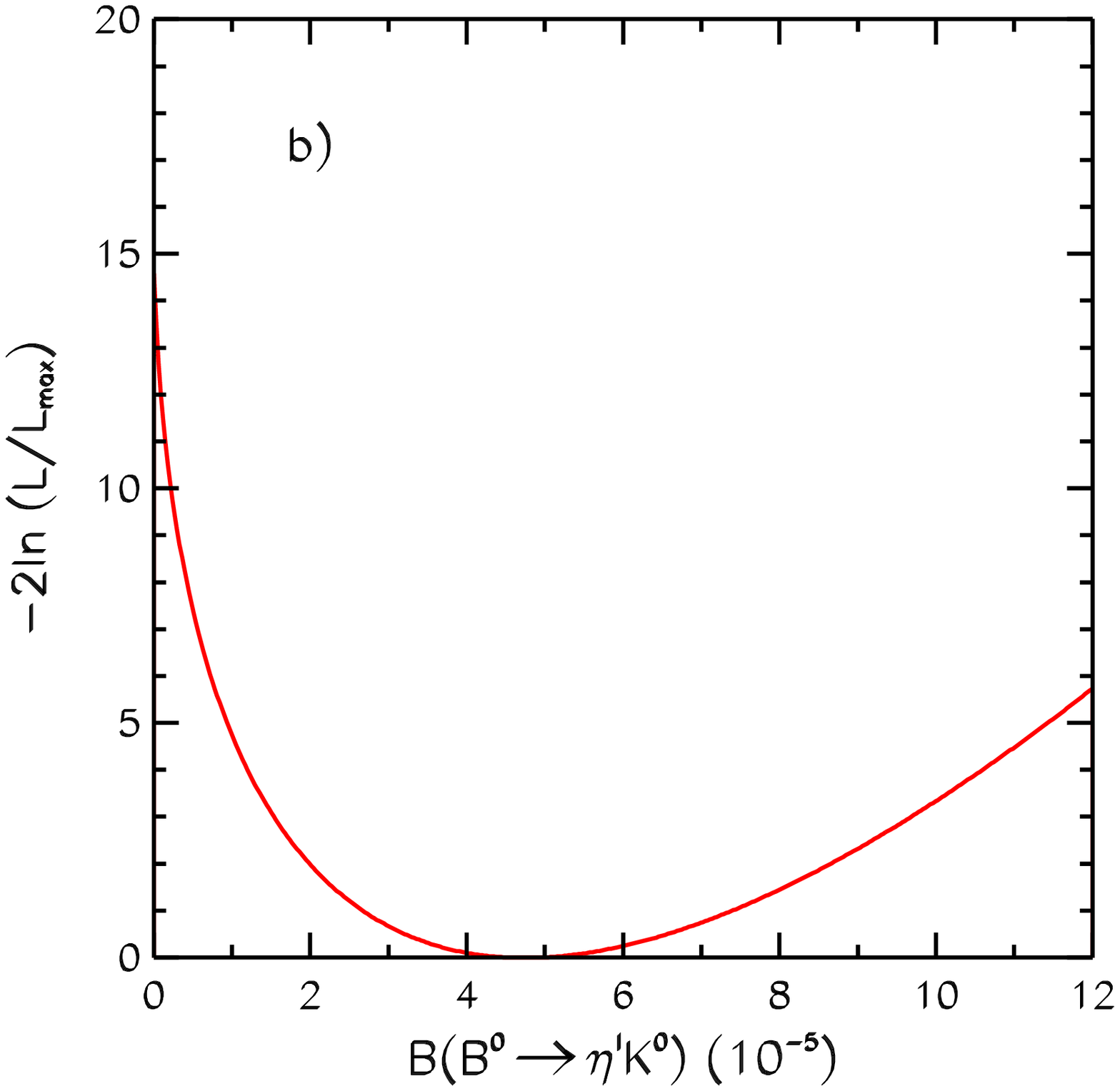}{1.0}
\caption{\label{contouretapr}%
(a) Likelihood function contours for $B^+\ra\eta^\prime h^+$; (b)
$-2\ln{\calL/\calL_{\rm max}}$ for \Betaprkz.}
\bigskip \bigskip
\psfiletwoBB{50 140 540 610}{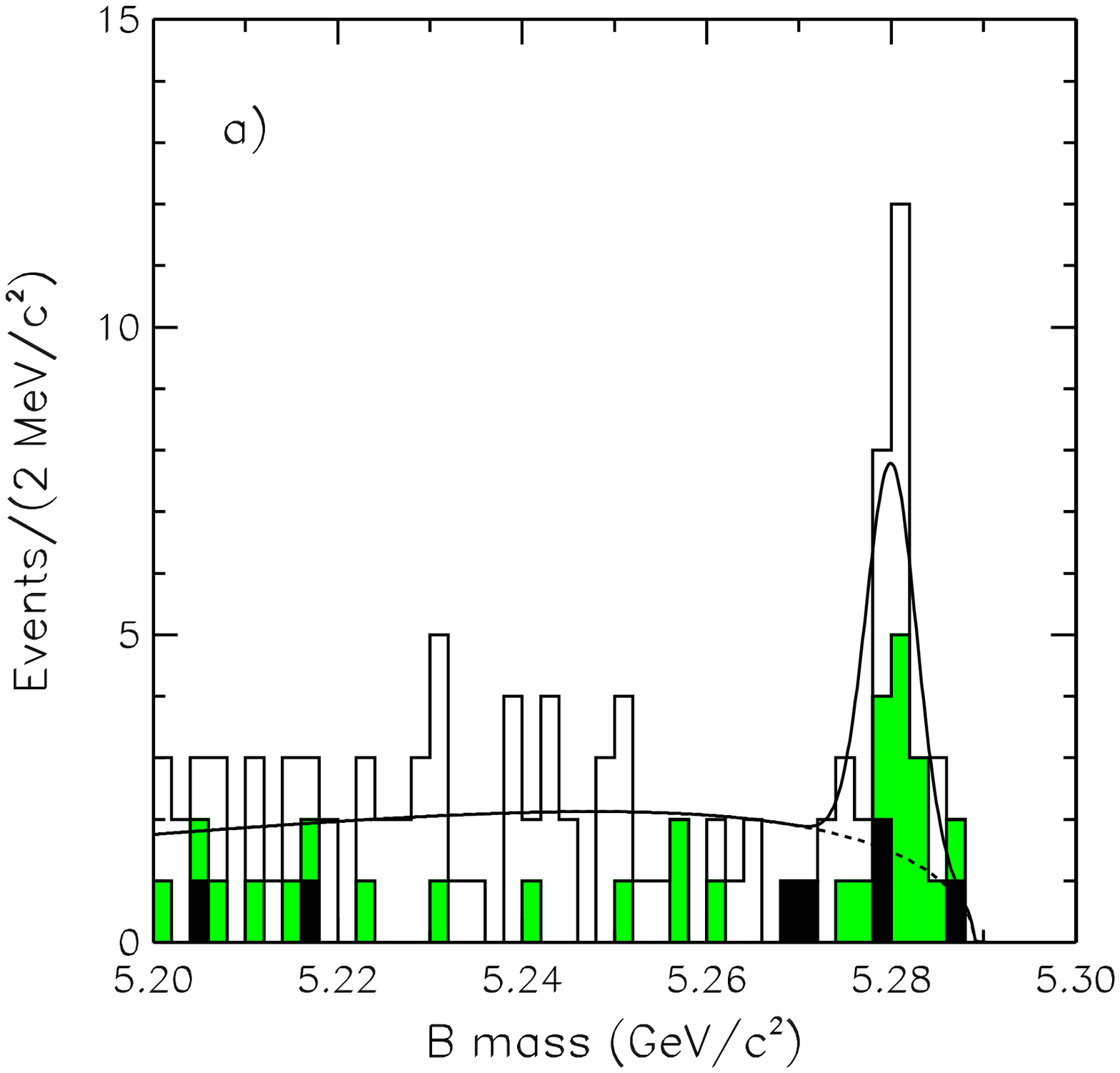}{50 140 540 610}{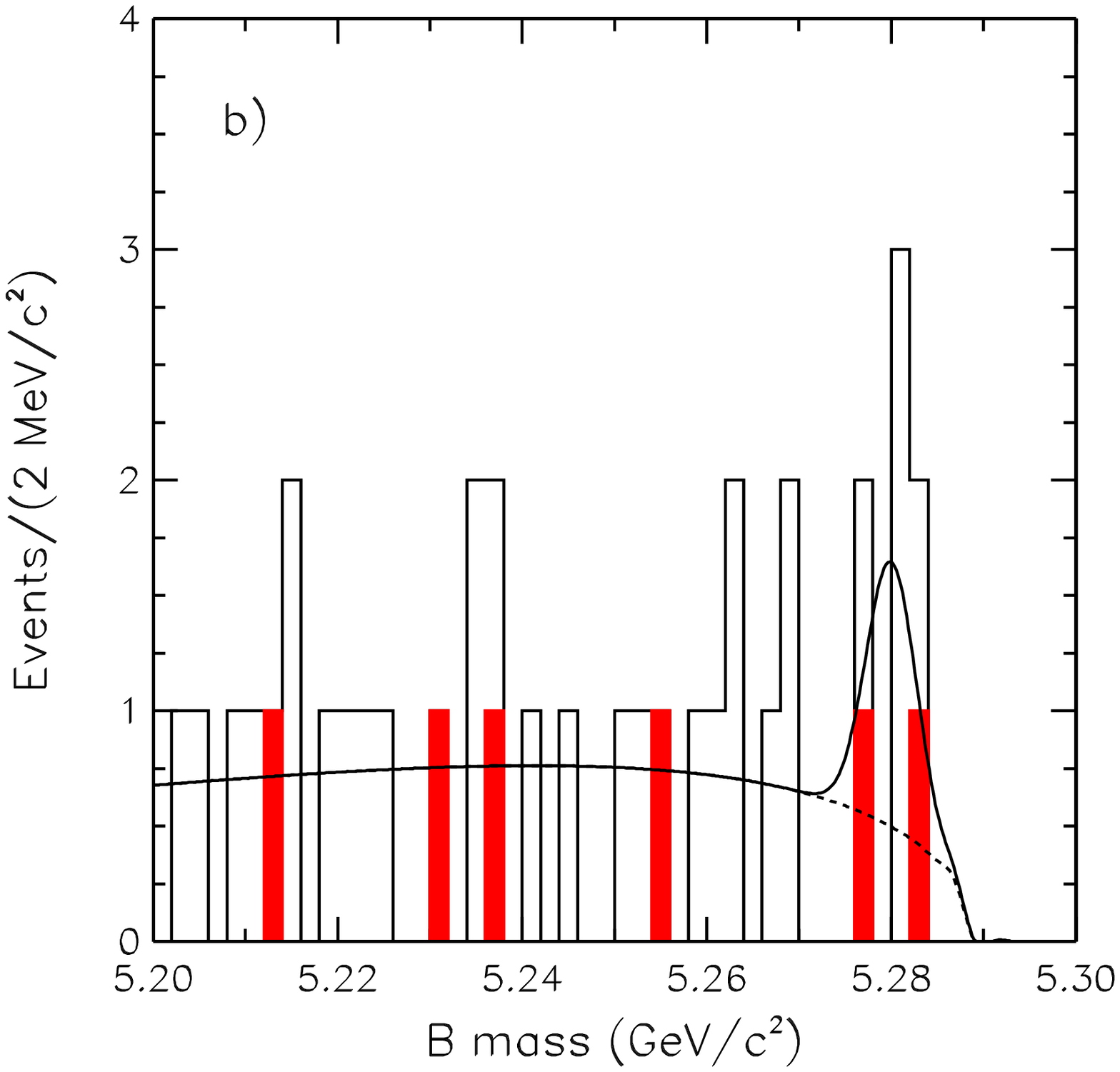}{1.0}
\caption{\label{etaprproj}%
Projections onto the variable $M$.  Overlaid on each plot as smooth
curves are the best fit functions (solid) and background components
(dashed), calculated with the variables not shown restricted to the
neighborhood
of expected signal.  The histograms show (a) $B^+\ra\eta^\prime
h^+$ with $\eta^\prime\ra \eta \pi\pi\ (\eta\ra\pi^+\pi^-\piz$, dark shaded),
$\eta^\prime\ra \eta \pi\pi\ (\eta\ra\gamma\gamma$, light shaded),
and $\eta^\prime\ra \rho\gamma$ (open); (b) \etaprkz\ with
$\eta^\prime\ra \eta \pi\pi$ (shaded) and $\eta^\prime\ra\rho\gamma$ (open).}
\end{figure}

The results of the ML fits for the $\etapr$ analyses are summarized in 
Table \ref{individtaba}.  A strong signal for \Betaprkp\ is found in both the 
\etaprepp\ ($5.2\sigma$) and \etaprrg\ ($4.8\sigma$) channels. 
Combining these with evidence from the chain \etaprepp, \etathreepi\
yields a significance of $7.5\sigma$ as shown in Fig.~\ref{contouretapr}a.
All significances given here and below include systematic errors in the
yield.  These are obtained from a Monte Carlo convolution of the likelihood 
function with resolution functions (assumed Gaussian) for the parameters, 
including their most important correlations. Efficiency systematics are
included as described above.  The combined significance for
the \Betaprkz\ decay is $3.8\sigma$ as shown in Fig.~\ref{contouretapr}b.
The projection plots for these signals are shown in Fig.~\ref{etaprproj}.

Similarly, the results for the ML fits for the $\eta$ final states are 
summarized in Table \ref{individtabb}.  Only limits are obtained for
these modes, though they are quite restrictive limits in many cases. 
For final states with multiple secondary channels, the value
of $-2\ln{\cal L}$ is summed for each branching fraction bin and
the final branching fraction or upper limit s extracted from the combined
distribution. Table~\ref{combtab} shows the final results for the $\etapr$ 
and $\eta$ decay modes, as well as previously published theoretical estimates.
Further details concerning the $\etapr$ and $\eta$ modes can be found in
reference \ref{etaprpub}.
\begin{table}[htbp]
\caption{Measurement results for $\eta$ decay modes.  Columns list the final 
states (with secondary decay modes as subscripts), event yield from the fit,
reconstruction efficiency $\epsilon$, total efficiency with secondary
branching fractions ${\cal B}_s$, and the resulting $B$ decay branching
fraction ${\cal B}$.  }
\vspace{0.3cm}
\begin{center}
\begin{tabular}{lcrrc}
\dbline
Final state&Fit events&$\epsilon$(\%)&$\epsilon\calB_s$(\%)&\calB($10^{-5})$\cr
\sgline
\etakgg       & $1.3^{+3.5}_{-1.3}$ &46&     17.9      & $<1.5$\cr
\etakthrp     & $0.0^{+2.5}_{-0.0}$ &28&      6.3      & $<3.1$\cr
\etakzgg      & $1.8^{+2.4}_{-1.6}$ &32&      4.2      & $<4.7$\cr
\etakzthrp    & $0.0^{+0.5}_{-0.0}$ &14&      1.1      & $<8.6$\cr
\etapigg      & $0.2^{+5.0}_{-0.2}$ &47&     18.2      & $<1.7$\cr
\etapithrp    & $0.0^{+1.8}_{-0.0}$ &29&      6.6      & $<2.6$\cr
\etapizgg     & $0.0^{+0.9}_{-0.0}$ &33&     13.0      & $<0.9$\cr
\etapizthrp   & $0.0^{+1.5}_{-0.0}$ &23&      5.5      & $<2.7$\cr
\etaetagg     & $1.1^{+1.7}_{-1.1}$ &34&      5.2      & $<3.0$\cr
\etaetathrp   & $0.0^{+1.3}_{-0.0}$ &24&      4.3      & $<2.9$\cr
\etaetasixp   & $0.0^{+0.5}_{-0.0}$ &16&      0.8      & $<9.8$\cr
\etakstpgg    & $0.7^{+3.6}_{-0.7}$ &25&      3.3      & $<8.8$\cr
\etakstpthrp  & $0.0^{+1.2}_{-0.0}$ &15&      1.2      & $<11.7$\cr
\etakstpggkz  & $0.0^{+1.2}_{-0.0}$ &24&      2.1      & $<5.7$\cr
\etakstpthrpkz& $0.0^{+1.0}_{-0.0}$ &14&      0.8      &$<16.0$\cr
\etakstzgg    & $5.2^{+4.0}_{-3.0}$ &32&      8.4      & $<4.6$\cr
\etakstzthrp  & $0.0^{+0.8}_{-0.0}$ &20&      3.1      & $<3.6$\cr
\etarhopgg    & $1.2^{+4.1}_{-1.2}$ &24&      9.9      & $<3.3$\cr
\etarhopthrp  & $2.5^{+4.1}_{-2.5}$ &14&      3.3      & $<11.2$\cr
\etarhozgg    & $0.2^{+4.0}_{-0.2}$ &36&     14.3      & $<1.9$\cr
\etarhozthrp  & $0.0^{+1.1}_{-0.0}$ &22&      5.1      & $<2.7$\cr
\dbline
\end{tabular}
\end{center}
\label{individtabb}
\end{table}

\begin{table}[ht]
\caption{Combined results for $\eta$ and $\etapr$ decay modes and expectations
from theoretical models.}
\vspace{0.4cm}
\def\notext{ & & \cr}
\begin{center}
\begin{tabular}{lccl}
\dbline
Decay mode  &    \calB($10^{-5})$      & Theory \calB($10^{-5}$)& References\cr
\sgline
\Betaprkp    & $6.5^{+1.5}_{-1.4}\pm0.9$&$0.7-4.1$&\cite{chau,kps,du}\cr
\Betaprkz    & $4.7^{+2.7}_{-2.0}\pm0.9$&$0.9-3.3$&\cite{chau,du}\cr
\Betaprpi    &          $<3.1$          &$0.8-3.5$&\cite{chau,kps,du}\cr
\Betaprpiz   &          $<1.1$          &$0.4-1.4$&\cite{chau,du}\cr
\Betapretapr &          $<4.7$          &$0.1-2.8$&\cite{chau,du}\cr
\Betapreta   &          $<2.7$          &$0.4-4.4$&\cite{chau,du}\cr
\Betaprkstp  &          $<13.$          &$0.1-0.9$&\cite{chau,kps,du}\cr
\Betaprkstz  &          $<3.9$          &$0.8-1.7$&\cite{chau,du}\cr
\Betaprrhop  &          $<4.7$          &$0.8-5.7$&\cite{chau,kps,du}\cr
\Betaprrhoz  &          $<2.3$          &$0.2-1.2$&\cite{chau,du}\cr
\Betak       &          $<1.4$          &$0.1-0.5$&\cite{chau,kps,du}\cr
\Betakz      &          $<3.3$          &$0.1-0.2$&\cite{chau,dean,du}\cr
\Betapi      &          $<1.5$          &$0.2-0.8$&\cite{chau,dean,kps,du}\cr
\Betapiz     &          $<0.8$          &$0.2-0.4$&\cite{chau,du}\cr
\Betaeta     &          $<1.8$          &$0.1-1.4$&\cite{chau,dean,du}\cr
\Betakstp    &          $<3.0$          &$0.1-1.3$&\cite{chau,kps,du}\cr
\Betakstz    &          $<3.0$          &$0.1-0.5$&\cite{chau,dean,du}\cr
\Betarhop    &          $<3.2$          &$0.7-4.4$&\cite{chau,dean,kps,du}\cr
\Betarhoz    &          $<1.3$          &$0.1-0.8$&\cite{chau,dean,du}\cr
\dbline
\end{tabular}
\end{center}
\label{combtab}
\end{table}

\begin{figure}[htbp]
\psfiletwoBB{60 150 530 610}{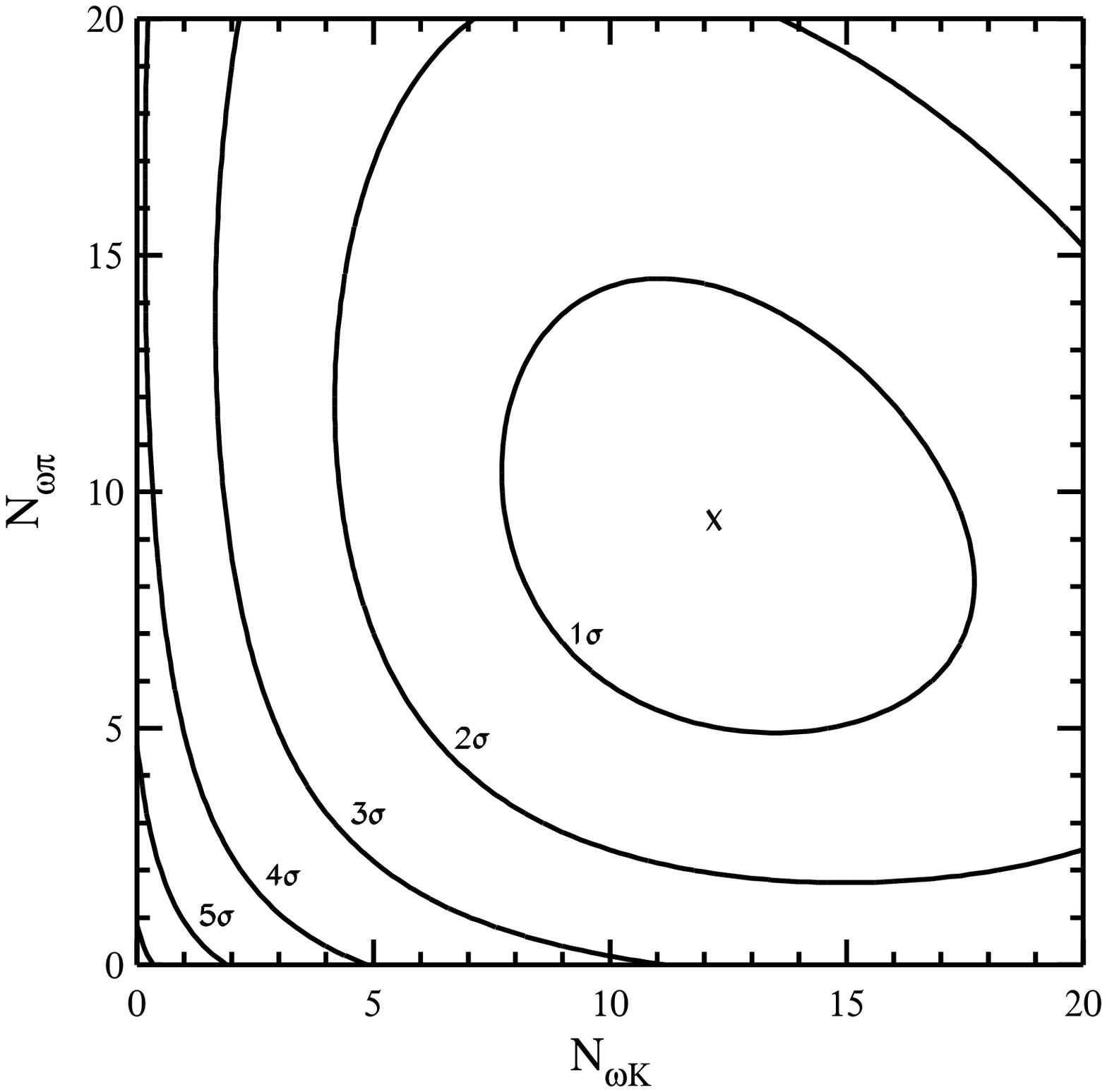}%
{25 0 520 490}{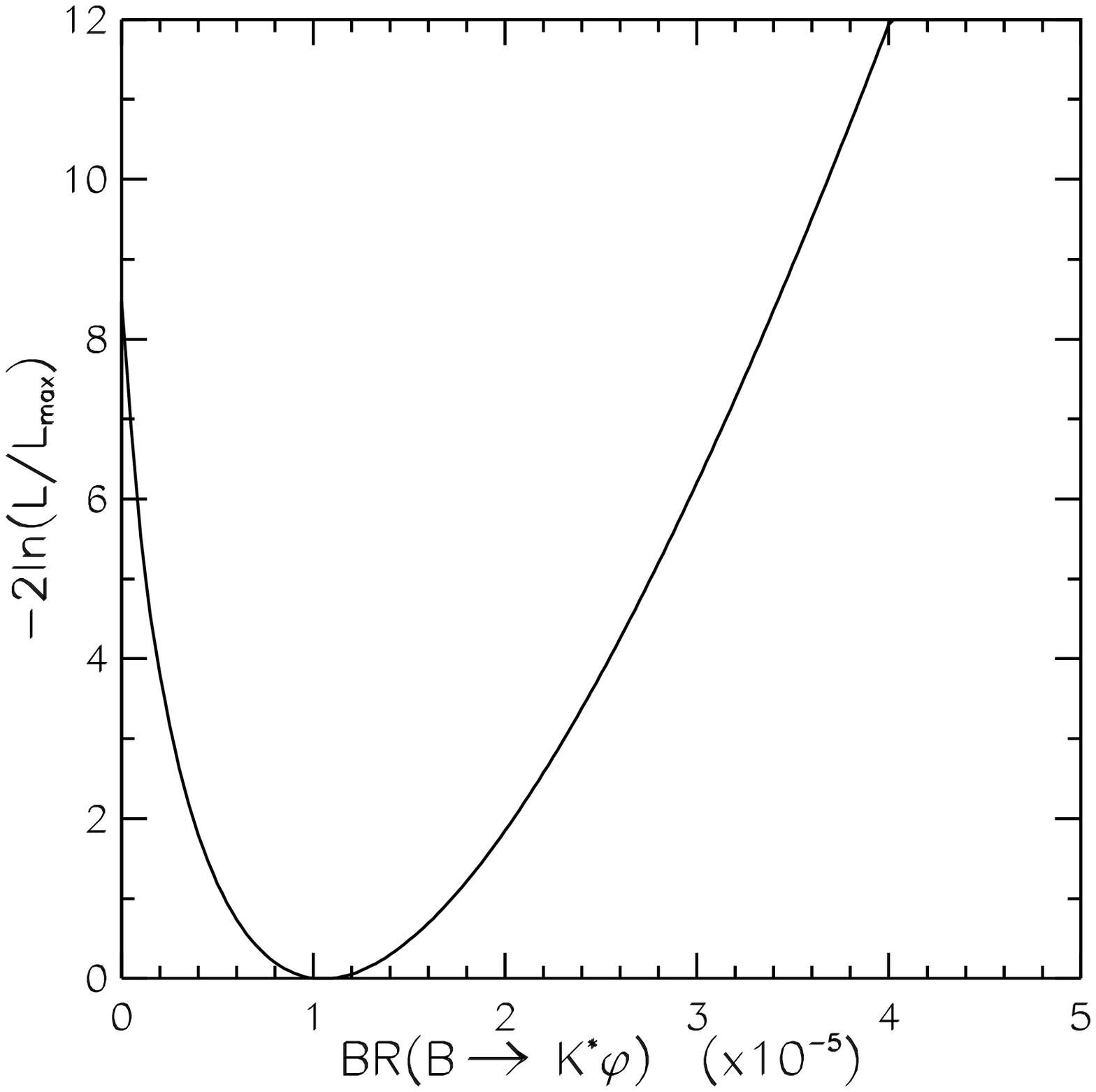}{1.0}
\caption{\label{contouromega}%
(a) Likelihood function contours for $B^+\ra\omega h^+$; (b) 
$-2\ln{\calL/\calL_{\rm max}}$ for $B\ra\phi K^*$.}
\bigskip\bigskip
\psfiletwoBB{50 140 540 610}{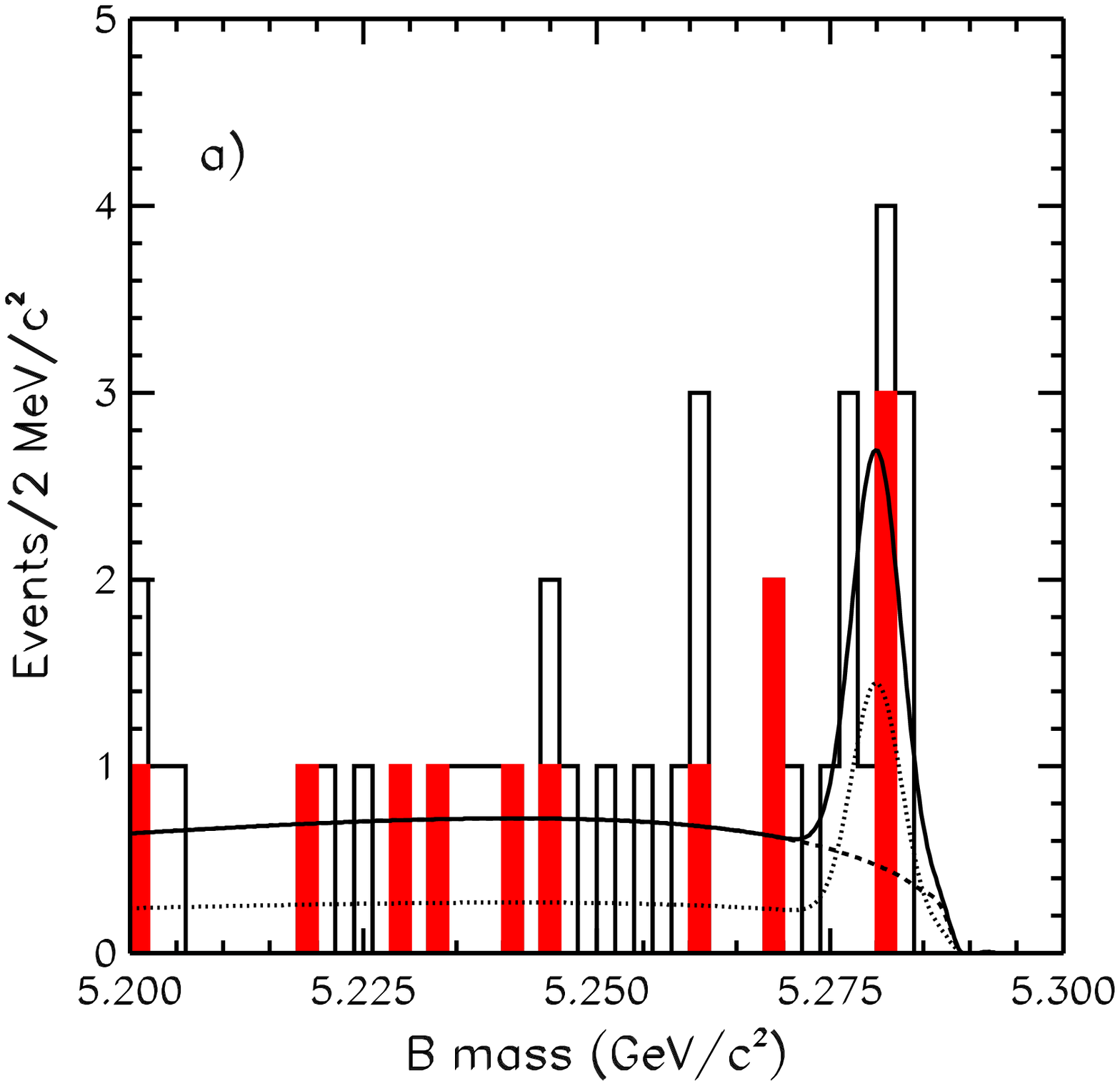}%
{25 0 540 490}{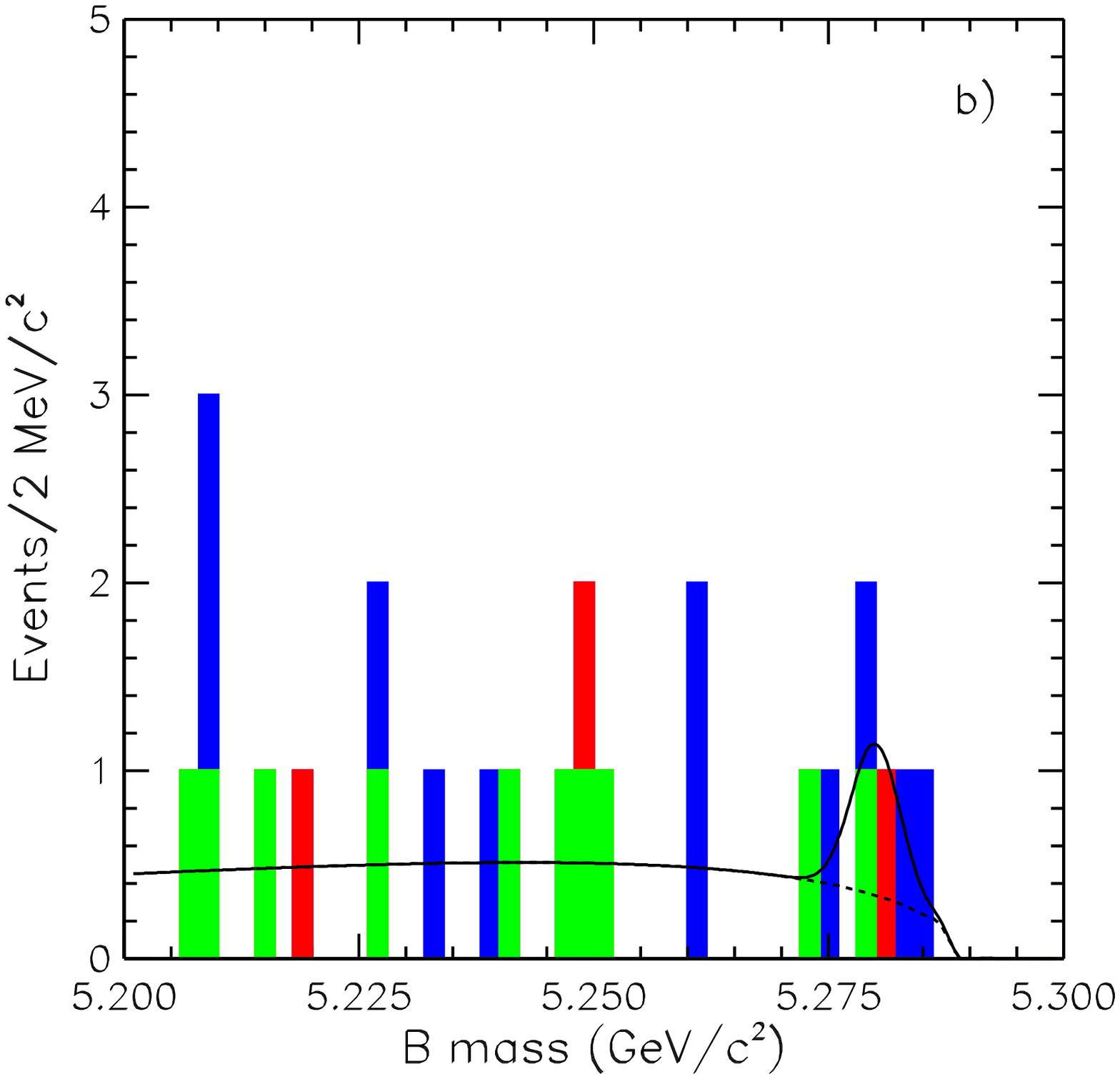}{1.0}
\caption{\label{projomega}%
Projection onto the variable $M$ for (a) \Bomegak (shaded) and \Bomegapi
(open) and (b) \Bphikst: \Bphikstz\ (dark shading), \Bphikstpkz, medium
shading), and \Bphikstpd, light shading). The solid line shows the result of 
the likelihood
fit, scaled to take into account the cuts applied to variables not
shown. The dashed line shows the background component, and in (a) the dotted
line shows the \Bomegak\ component of the fit only.}
\end{figure}

Finally, we give in Table \ref{omindivtab} results for the ML fits for
the $\omega$ and $\phi$ final states.  
Table \ref{omcombtab} provides a summary of these results,
where modes with multiple secondary channels have been combined.
A signal with $3.9\sigma$
significance is found for \Bomegak\ as shown in Fig.~\ref{contouromega}a.
The corresponding projection plot is shown in Fig.~\ref{projomega}a.
The significance for the combination of the $\phi K^{*+}$ and $\phi K^{*0}$
final states is marginal - $2.9\sigma$.  It is sensible to combine
these modes since the penguin diagrams for the $B^+$ and $B^0$ decays are
identical except for the spectator quark and all other processes are expected
to be negligible for these decays.  If the observed yield is interpreted
as a signal, a branching fraction of $(1.1^{+0.6}_{-0.5}\pm0.2)\times10^{-5}$
is obtained.  The plot of $-2\ln{\cal L}$ and fit 
projections are shown in Figs.~\ref{contouromega}b and \ref{projomega}b,
respectively.  
Further details concerning the $\omega$ and $\phi$ modes can be found in
reference \ref{omegapub}.

\begin{table}[htbp]
\caption{Measurement results for $\omega$ and $\phi$ decay modes.  
Columns list the final 
states (with secondary decay modes as subscripts), event yield from the fit,
reconstruction efficiency $\epsilon$, total efficiency with secondary
branching fractions ${\cal B}_s$, and the resulting $B$ decay branching
fraction ${\cal B}$.  }
\vspace{0.4cm}
\def\notext{ & & & & \cr}
\begin{center}
\begin{tabular}{lcrrc}
\dbline
Final state&Fit events&$\epsilon$(\%)&$\epsilon\calB_s$(\%)&\calB($10^{-5})$\cr
\sgline
\omegak       &$12.2^{+5.5}_{-4.5}$&28&25.1&$1.5^{+0.7}_{-0.6}\pm0.2$\cr
\omegakz      & $2.3^{+2.4}_{-1.5}$&15&      4.4      &$<5.7$  \cr
\omegapi      & $9.2^{+5.3}_{-4.3}$&29&     25.8      &$<2.3$  \cr
\omegah       & $21.4^{+6.5}_{-5.6}$&29&25.5&$2.5^{+0.8}_{-0.7}\pm0.3$\cr
\omegapiz     & $2.4^{+2.9}_{-1.8}$&24&     20.9      &$<1.4$  \cr
\omegaetaprd  & $0.1^{+1.9}_{-0.1}$&16&      2.4      &$<6.4$  \cr
\omegaetaprrg & $5.1^{+3.6}_{-2.7}$&16&      4.2      &$<9.2$  \cr
\omegaetagg   & $0.0^{+1.5}_{-0.0}$&24&      8.5      &$<2.0$  \cr
\omegaetathrp & $0.0^{+0.5}_{-0.0}$&15&      3.2      &$<2.8$  \cr
\omegakstpd   & $1.1^{+2.6}_{-1.1}$& 7&      2.0      &$<12.9$  \cr
\omegakstpkz  & $4.5^{+3.6}_{-2.8}$&16&      3.2      &$<10.9$ \cr
\omegakstzd   & $2.1^{+3.6}_{-2.1}$&22&     13.1      &$<2.3$  \cr
\omegarhop    & $2.5^{+4.4}_{-2.5}$& 8&      6.8      &$<6.1$  \cr
\omegarhoz    & $0.0^{+1.7}_{-0.0}$&24&     21.1      &$<1.1$  \cr
\omegaomega   & $0.3^{+2.6}_{-0.3}$&15&     11.9      &$<1.9$  \cr
\phik         & $0.0^{+0.8}_{-0.0}$&47&     23.1      &$<0.5$  \cr
\phikz        & $1.9^{+2.0}_{-1.2}$&32&      5.3      &$<3.1$  \cr
\phipi        & $0.0^{+0.9}_{-0.0}$&49&     24.0      &$<0.5$  \cr
\phipiz       & $0.0^{+0.6}_{-0.0}$&31&     15.1      &$<0.5$  \cr
\phietaprd    & $0.0^{+0.5}_{-0.0}$&26&      2.2      &$<3.5$  \cr
\phietaprrg   & $2.7^{+3.1}_{-2.1}$&30&      4.4      &$<6.3$  \cr
\phietagg     & $0.0^{+0.6}_{-0.0}$&39&      7.5      &$<1.3$  \cr
\phietathrp   & $0.0^{+0.5}_{-0.0}$&24&      2.7      &$<2.9$  \cr
\phikstpd     & $2.6^{+3.3}_{-2.4}$&26&      4.4      &$<5.6$  \cr
\phikstpkz    & $1.7^{+2.0}_{-1.1}$&29&      3.4      &$<5.3$  \cr
\phikstzd     & $3.2^{+3.2}_{-2.1}$&39&     12.7      &$<2.2$  \cr
\phikstzkz    & $0.0^{+1.9}_{-0.0}$&18&      1.0      &$<8.0$  \cr
\phirhop      & $0.0^{+2.3}_{-0.0}$&34&     16.7      &$<1.6$  \cr
\phirhoz      & $0.8^{+4.4}_{-0.8}$&41&     20.0      &$<1.3$  \cr
\phiomega     & $0.8^{+2.5}_{-0.8}$&23&     10.2      &$<2.1$  \cr
\phiphi       & $0.4^{+1.4}_{-0.4}$&40&      9.7      &$<1.2$  \cr
\dbline
\end{tabular}
\end{center}
\label{omindivtab}
\end{table}

\begin{table}[ht]
\caption{Combined results for $\omega$ and $\phi$ decay modes and expectations
from theoretical models.}
\vspace{0.4cm}
\def\notext{ & & \cr}
\begin{center}
\begin{tabular}{lccl}
\dbline
Decay mode  &    \calB($10^{-5})$     & Theory \calB\ ($10^{-5}$)&References\cr
\sgline
\Bomegak    &$1.5^{+0.7}_{-0.6}\pm0.2$&$0.1-0.7$&\cite{chau,dean,kps,du} \cr
\Bomegakz   &          $<5.7$         &$0.1-0.4$&\cite{chau,dean,du} \cr
\Bomegapi   &          $<2.3$         &$0.1-0.7$&\cite{chau,dean,kps,du} \cr
\Bomegah    &$2.5^{+0.8}_{-0.7}\pm0.3$& -       & - \cr
\Bomegapiz  &          $<1.4$         &$0.01-1.2$&\cite{chau,dean,du} \cr
\Bomegaetapr&          $<6.0$         &$0.3-1.7$&\cite{chau,du} \cr
\Bomegaeta  &          $<1.2$         &$0.1-0.5$&\cite{chau,du} \cr
\Bomegakstp &          $<8.7$         &$0.04-1.5$&\cite{chau,dean,kpsvv} \cr
\Bomegakstz &          $<2.3$         &$0.2-0.8$&\cite{chau,dean} \cr
\Bomegarhop &          $<6.1$         &$1.0-2.5$&\cite{chau,dean,kpsvv} \cr
\Bomegarhoz &          $<1.1$         &$0.04$   &\cite{chau} \cr
\Bomegaomega&          $<1.9$         &$0.04-0.3$&\cite{chau,dean} \cr
\Bphik      &          $<0.5$         &$0.07-1.6$&\cite{desh,chau,dean,fl,dav,kps,du}\cr
\Bphikz     &          $<3.1$         &$0.07-1.3$&\cite{desh,chau,dean,fl,dav,du}\cr
\Bphipi     &          $<0.5$         & $<<0.1$ &\cite{xing,dean,fl,kps,du}\cr
\Bphipiz    &          $<0.5$         & $<<0.1$ &\cite{xing,dean,fl,du}\cr
\Bphietapr  &          $<3.1$         & $<<0.1$ &\cite{xing,du}  \cr
\Bphieta    &          $<0.9$         & $<<0.1$ &\cite{xing,dean,du}  \cr
\Bphikstp   &          $<4.1$         &$0.02-3.1$&\cite{desh,chau,dean,dav,kpsvv}\cr
\Bphikstz   &          $<2.1$         &$0.02-3.1$&\cite{desh,chau,dean,dav} \cr
\Bphirhop   &          $<1.6$         & $<<0.1$ &\cite{xing,dean,kpsvv} \cr
\Bphirhoz   &          $<1.3$         & $<<0.1$ &\cite{xing,dean} \cr
\Bphiomega  &          $<2.1$         & $<<0.1$ &\cite{xing,dean} \cr
\Bphiphi    &          $<1.2$         &none & \cr
\dbline
\end{tabular}
\end{center}
\label{omcombtab}
\end{table}

\section{Evidence for the inclusive decay \etaprinc\ from CLEO}

Evidence also has been found for the inclusive decay \etaprinc.  In this
analysis the state \xs\ is defined as a charged kaon accompanied by from
zero to four pions, of which at most one can be a $\piz$.  The momentum of 
$\etapr$ mesons,
reconstructed with the decay chain \etaprepp, \etatogg, is required to be
in the range $2.0<p_\etapr<2.7$ GeV/c in order to reduce background
from $b\ra c$ processes.  The values of \DE\ and $M$, as defined
above, are required to satisfy $|\DE|<0.1$ GeV and $M>5.275$ GeV.

\displaytwo{htbp}{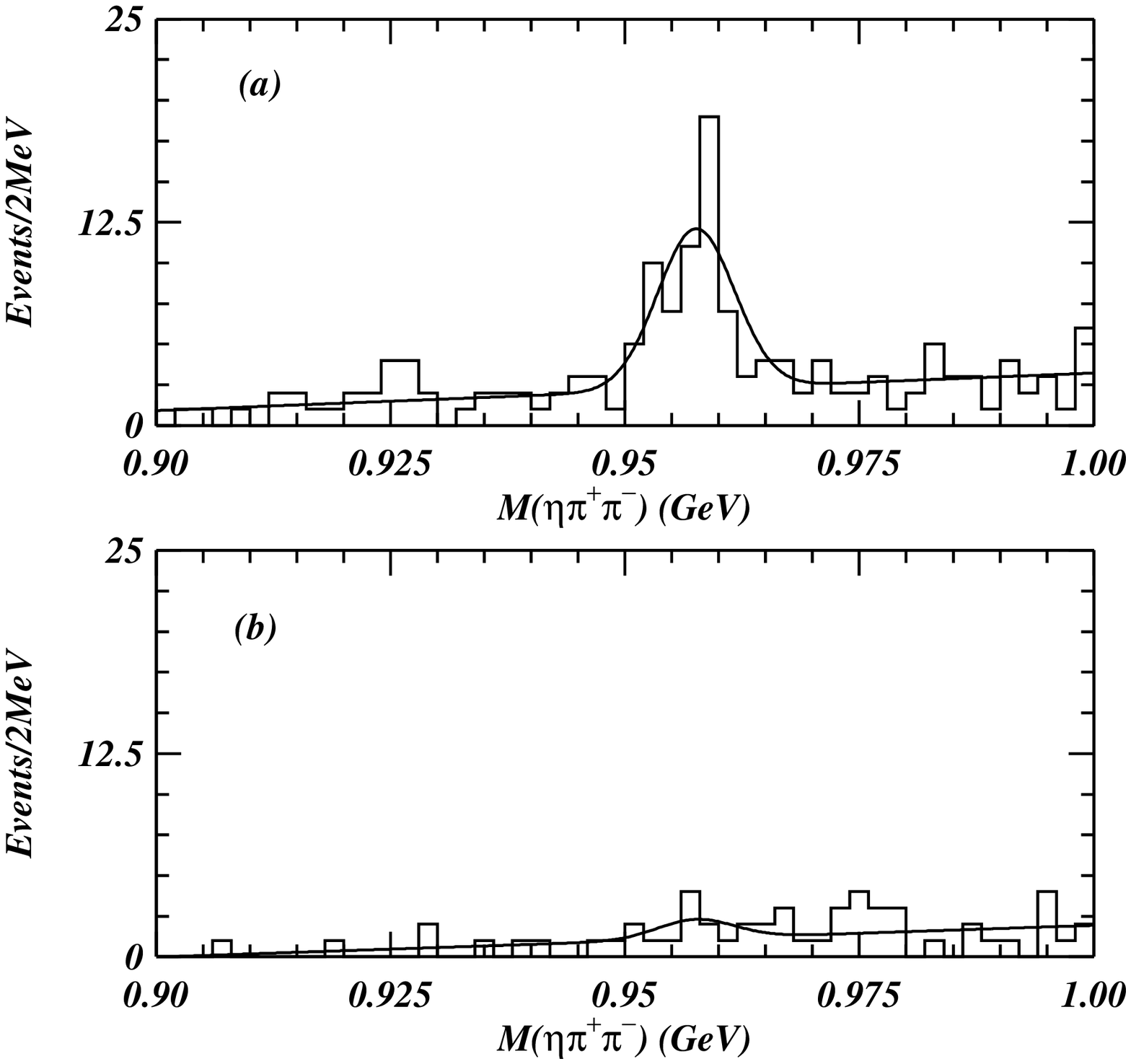}{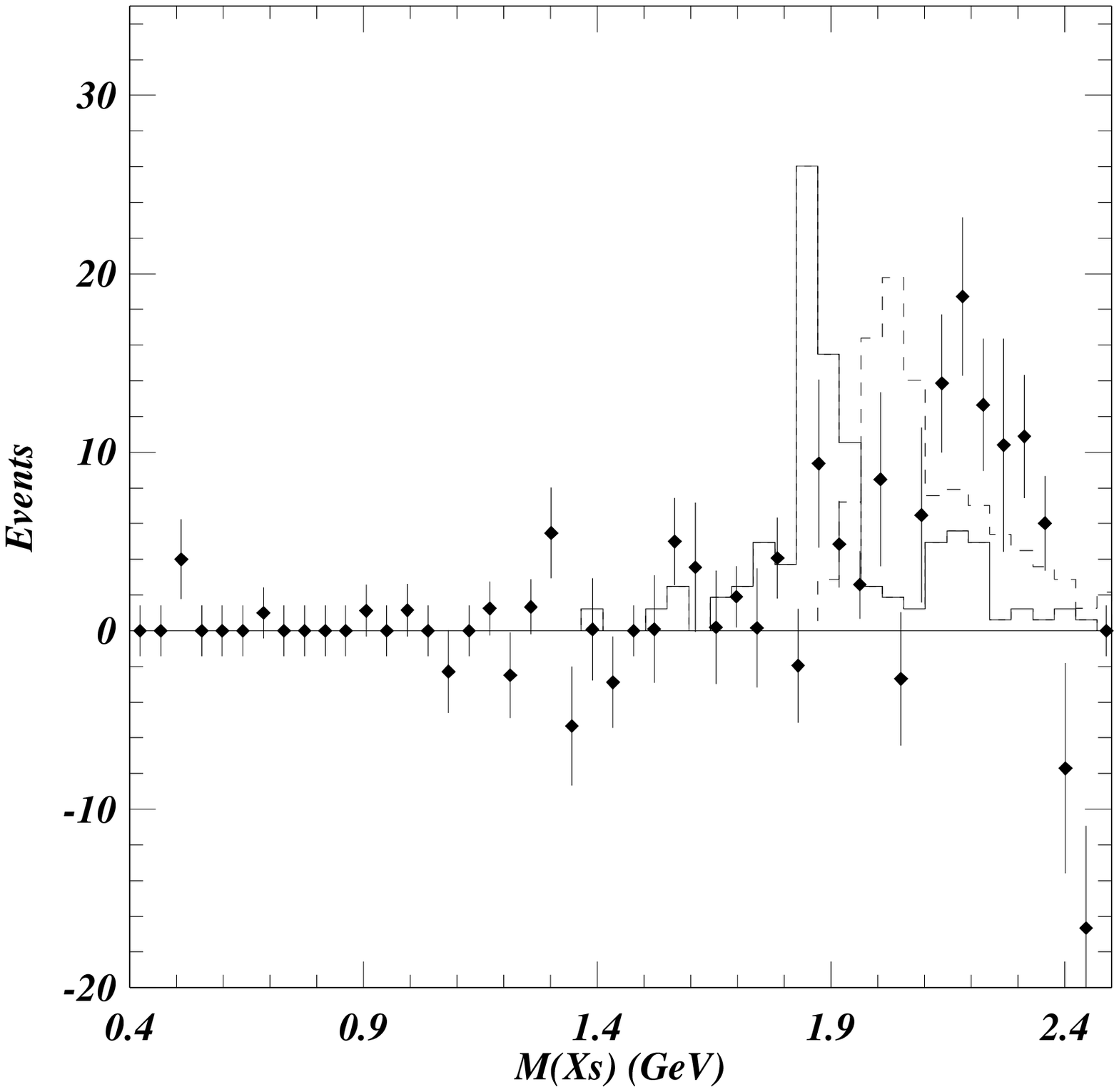}
{The $\etapr$ mass distribution for (a) on-resonance and (b) off-resonance
data.}{Distribution of \xs\ mass for data
(points with error bars) and possible backgrounds: $B\ra D\etapr$ (solid
histogram) and $B\ra D^*\etapr$ (dashed histogram). The normalization of
the backgrounds is arbitrary.}

The $\etapr$ mass distribution is shown in Fig. \ref{etapinc_mass}; a 
clear signal of $39\pm10$ events is seen for on-resonance data and none for 
the off-resonance sample. The signal,
obtained by subtracting the off-resonance data
in bins of \xs\ mass, is plotted in Fig. \ref{xsmass_cbg}.
Note the four events corresponding to \Betaprkp\ and the absence of
events in the $K^*(892)$ mass region, both consistent with the exclusive
results given above.  Also shown in Fig. \ref{xsmass_cbg} 
are distributions for potential background modes such as
$B\ra D\etapr$ and $B\ra D^*\etapr$.  Though these also tend to have large
\xs\ mass, they are more peaked than the data.  These and other
studies suggest that the observed signal does not arise primarily from
color-suppressed $b\ra c$ decays, though it is difficult to rule this
out completely without better models of such processes.  The efficiency
is calculated assuming that the signal arises solely from gluonic penguin
decays, with an equal admixture of \xs\ states from the kaon mass up to
$K_4^*(2200)$.  The efficiency of ($5.5\pm0.3$)\% leads to
$\calB(\etaprinc)=(6.2\pm1.6\pm1.3)\times10^{-4}$ for $2.0<p_\etapr<2.7$ GeV/c.
The systematic error is dominated by the uncertainty in the \xs\ modelling.

\section{Conclusion}

CLEO has observed for the first time five charmless hadronic $B$ decay
modes.  The measured branching fractions range from (1--7)$\times10^{-5}$.
All of these modes involve $K$ mesons while none of the related
modes involving pions have yet been observed.  This suggests 
that penguin loop diagrams are playing a dominant role in these decays.
These new results have sparked a considerable amount of theoretical
activity during 1997.  Fleischer and Mannel \cite{flman}
claim that the fact that the
ratio of the $K^\pm\pi^\mp$ and $K^0\pi^\pm$ modes is less than one (with
large errors) may soon facilitate useful bounds on the CKM angle $\gamma$.
Many recent papers point out that rescattering effects and electroweak 
penguins may complicate or invalidate this method.  

The branching fraction for \Betaprkp, $(6.5^{+1.5}_{-1.4}\pm0.9)\times10^{-5}$,
is several times larger than other charmless hadronic $B$ decays.  This was
unexpected, though it had been pointed out by Lipkin \cite{lipkin} that 
interference effects between the two penguin diagrams, Fig.~\ref{feynfig}e and
\ref{feynfig}f, enhance the \etaprk\ rate and suppress \etak.  The
branching fractions and upper limits given in Table~\ref{combtab} clearly 
exhibit this pattern.  There have been a variety of recent
effective-Hamiltonian calculations \cite{aliGreub}\ which try to account
for processes measured here, the large rate for \Betaprk\ in particular.
They generally employ spectator and factorization \cite{BSW} approximations,
though the validity of the latter has been established only in $b\ra c$
processes.  These calculations have suggested enhancements from larger 
form factors \cite{datta,kagan}, smaller strange-quark mass \cite{kagan},
and variation of the effective number of colors \cite{aliGreub,oh}.
Others have suggested a contribution from the QCD gluon anomaly (Fig.\
\ref{feynfig}h) or other flavor singlet processes in constructive 
interference with the penguins \cite{soni,etaCP,hairpin,cheng,zhitnitsky}.
Given the experimental errors, most of these calculations can account
for the data, though the models with additional singlet contributions
appear to be needed unless the branching fraction for \Betaprkp\ is
reduced substantially when further data is obtained.

The theoretical situation with the $\omega$ and $\phi$ modes is also
quite interesting.  We also establish 90\% CL lower limits on the
branching fractions \Bomegak\ and \Bomegah\ of $8.4\times10^{-6}$ and 
$1.6\times10^{-5}$, respectively.  Predictions for these rates 
tend to be smaller than the observed rate for most values of the color
parameter $\xi$ \cite{aliGreub,oh,cheng}.  Predictions for \Bphik\ tend
to be larger than the limited presented here; the combination of these
upper and lower limits rules out all values of $\xi$ {\it for these models}
at $>90$\% CL, though additional variation of theoretical parameters
could probably account for the data.  A recent calculation \cite{ciuchini},
involving an enhanced contribution from charmed quarks in the penguin
loop, also has difficulty accounting for a large rate in the \omegak\
channel but predicts large branching fractions for final states such
as $\omega K^*$ and $\phi K^*$.

There have also been many attempts to explain the even more surprising
excess of $\etapr$ inclusive events.  Atwood and Soni \cite{soni} first 
suggested an enhancement in this process arising from the anomalous coupling of
gluons with the $\etapr$ meson, analogous to the exclusive diagram shown
in Fig. \ref{feynfig}h.  Other
authors \cite{datta,kagan,hou,chao} have considered this process, though
without a consensus whether the anomaly can account for the inclusive rate.
Prospects are excellent for resolution of many of these 
issues during 1998 as new data become available.

\section*{Acknowledgments}
I'd like to thank the organizers for an enjoyable and stimulating conference.  
I thank my CLEO colleagues for their assistance and helpful discussions, 
especially Bruce Behrens, Tom Browder, Jim Fast, Bill Ford, Andrei Gritsan, 
and Jean Roy.
I also gratefully acknowledge many useful discussions with A. Ali, H-Y.
Cheng, A. Datta, T. DeGrand, A. Kagan, H. Lipkin, S. Oh, and A. Soni.
This work was supported by the Department of Energy under
grant DE-FG02-91ER40672.


\begin{thebibliography}{99}
\bibitem{mlfit} The $K\pi$ and $\pi\pi$ analyses use a similar (not
``extended") ML fit in which event fractions rather than the total
numbers of events are used.
\end{thebibliography}
\end{document}